\documentclass[aps,prd,twocolumn,groupedaddress,floatfix,longbibliography,superscriptaddress]{revtex4-1}

\usepackage{amsmath}
\usepackage{amssymb}
\usepackage{amsfonts}
\usepackage{bbold}
\usepackage{bm}
\usepackage{cancel}
\usepackage{times,float}
\usepackage{graphicx}
\usepackage[usenames,dvipsnames,svgnames]{xcolor}
\usepackage{hyperref}
\hypersetup{colorlinks=true, linkcolor=Blue, citecolor=Blue,urlcolor=Blue}
\usepackage{multirow}
\usepackage{ulem}
\usepackage{color,epsfig}
\usepackage{slashed}
\usepackage{feynmf}
\usepackage{array}
\usepackage{physics}
\usepackage{subcaption}
\usepackage{soul}

\begin{document}


\title{Planar Hall effect in Weyl semimetals induced by pseudoelectromagnetic fields}

\author{L. Medel Onofre}
\email{leonardo.medel@correo.nucleares.unam.mx}
\address{Instituto de Ciencias Nucleares, Universidad Nacional Aut\'{o}noma de M\'{e}xico, 04510 Ciudad de M\'{e}xico, M\'{e}xico}

\author{A. Mart\'{i}n-Ruiz}
\email{alberto.martin@nucleares.unam.mx}
\address{Instituto de Ciencias Nucleares, Universidad Nacional Aut\'{o}noma de M\'{e}xico, 04510 Ciudad de M\'{e}xico, M\'{e}xico}

\begin{abstract}
The planar Hall effect (PHE), the appearance of an in-plane transverse voltage in the presence of coplanar electric and magnetic fields, has been ascribed to the chiral anomaly and Berry curvature effects in Weyl semimetals. In the presence of position- and time-dependent perturbations, such as strain, Weyl semimetals react as if they would be subjected to emergent electromagnetic fields, kwnon as pseudo-fields. In this paper we investigate the possibility of inducing nonlinear phenomena, including the PHE, in strained Weyl semimetals. Using the chiral kinetic theory in the presence of pseudo-fields, we derive general expressions for the magnetoconductivity tensor by considering the simultaneous effects of the Berry curvature and orbital magnetic moment of carriers, which are indeed of the same order of magnitude.  Since pseudo-fields couple to the Weyl fermions of opposite chirality with opposite signs, we study chirality-dependent phenomena, including the longitudinal magnetoconductivity and the planar Hall effect. We discuss our results in terms of the chiral anomaly with pseudo-fields. These may open new possibilities in chiralitytronics.
\end{abstract}

\maketitle


\section{Introduction}

Weyl semimetals (WSMs) are topologically nontrivial conductors in which the non-degenerate valence and conduction bands touch at isolated points (the so called Weyl nodes) in the Brillouin zone \cite{RevModPhys.90.015001}. Near these touching-points, the electron spectrum can be described by the Weyl equation, originally introduced in the particle physics context. The Weyl nodes occur in pairs of opposite chirality which act as a source and sink of Berry curvature in reciprocal space, and the WSM phase is topologically protected by a nonzero Berry flux across the Fermi surface.

A distinguishing transport property of WSMs with broken time-reversal symmetry is the anomalous Hall effect, which arises when the conduction (valence) band is completely empty (filled) \cite{PhysRevB.86.115133}. Another intriguing property of WSMs is the chiral anomaly, i.e., the nonconservation of the chiral current in the presence of parallel electric and magnetic fields: $\partial _{\mu} J ^{\mu} _{5} = \frac{e^{2}}{2 \pi ^{2} \hbar ^{2}} \boldsymbol{E} \cdot \boldsymbol{B}$. The appearance of a positive longitudinal magnetoconductance has been regarded as a manifestation of the chiral anomaly \cite{PhysRevB.88.104412, PhysRevLett.113.247203}. However, the magnetoconductivity tensor receives additional contributions, not related with the chiral anomaly, which indeed reverses the overall sign of the magnetoconductance.

The planar Hall effect (PHE), the appearance of an in-plane transverse voltage in the presence of coplanar electric and magnetic fields, has been ascribed to the chiral anomaly in WSMs as well as to Berry curvature effects \cite{PhysRevLett.119.176804, PhysRevB.100.115139, PhysRevB.102.121105, PhysRevB.99.115121, PhysRevB.102.205107, Nag_2021, YADAV2022115444}. However, as in the case of the longitudinal magnetoconductance, within the semiclassical Boltzmann transport theory, the PHE is not described solely by the Berry curvature. In fact, as we show in this paper, the orbital magnetic moment (OMM) of charge carriers contributes also to the PHE, in much the same order of magnitude than the Berry curvature contribution, and therefore it cannot be disregarded. In a similar footing, the electrochemical transport in WSMs has been associated with the chiral anomaly, with the orbital magnetic moment playing a fundamental role, since the statistical transport is sensitive to the spatial gradients of the distribution function \cite{PhysRevB.103.035102}.

In conventional transport experiments, Weyl quasiparticles are coupled to the electromagnetic fields $\boldsymbol{E}$ and $\boldsymbol{B}$, which cannot be used as probe for chirality ($\chi = \pm 1 $) since they do not differentiate the nodes. However, an interesting phenomena arising in Dirac matter is the fact that elastic deformations of the lattice couple to the electronic Hamiltonian as pseudo-electromagnetic gauge potentials $\tilde{\boldsymbol{A}} _{\chi} ^{\mbox{\scriptsize el} } $ and $\tilde{\Phi} _{\chi} ^{\mbox{\scriptsize el} } $, which define pseudo-electromagnetic fields as usual, $\boldsymbol{E} _{\chi } ^{\mbox{\scriptsize el} }  = - \nabla \Phi _{\chi} ^{\mbox{\scriptsize el} }  - \partial _{t} \boldsymbol{A} _{\chi} ^{\mbox{\scriptsize el} } $ and $\boldsymbol{B} _{\chi} ^{\mbox{\scriptsize el} }  = \nabla \times \boldsymbol{A} _{\chi} ^{\mbox{\scriptsize el} } $, known as pseudo-fields or elastic-fields \cite{Ilan2020}.  These pseudo-fields can be expressed as the sum of two terms: $\boldsymbol{E} _{\chi } ^{\mbox{\scriptsize el} }  = \boldsymbol{\mathcal{E}} + \chi \boldsymbol{E} _{5} $ and $\boldsymbol{B} _{\chi } ^{\mbox{\scriptsize el} }  =  \boldsymbol{\mathcal{B}} + \chi \boldsymbol{B} _{5} $. While the $\boldsymbol{\mathcal{E}}$ and $\boldsymbol{\mathcal{B}}$ couple to the Weyl nodes in a similar fashion as electromagnetic fields do (with the same sign),  $\boldsymbol{E} _{5}$ and $\boldsymbol{B} _{5}$ couple to the nodes in an axial fasion (i.e. they couple opposite chiral fermions with opposite signs). The notation for  $\boldsymbol{E} _{5}$ and $\boldsymbol{B} _{5}$ in inherited from the high-energy physics literature, where axial fields couple to the Dirac matrix $\gamma _{5}$. In a strained material, the hoping parameters between atomic orbitals and on-site energies are both changed, and the modifications are driven by the components of the strain-tensor $u_{ij}$. Therefore, the axial fields become determined by the position- and/or time-dependence of the the strain tensor. For example, in the case of strained graphene, the induced pseudo-gauge fields couple to the Dirac fermions oppositely in the two valleys $\boldsymbol{K}$ and $\boldsymbol{K}'\,$ \cite{Guinea2010}. This gave rise to a new line of research called straintronics \cite{AMORIM20161, PhysRevB.65.235412}. More recently, the study of strain-induced gauge fields in Weyl semimetals has attracted great attention \cite{PhysRevLett.115.177202, PhysRevLett.115.177202},  since it could opens a pathway for a prolific industry associated with straintronics and chiraltronics.

In the case of WSMs, mechanical strain shifts the position of the Weyl nodes in momentum and/or energy, which can be effectively described in terms of pseudo-electromagnetic fields \cite{PhysRevLett.115.177202, PhysRevLett.115.177202}. This affects the low-energy description of Weyl fermions, which now have to include the pseudo-gauge potentials \cite{PhysRevB.87.235306, PhysRevB.89.081407, PhysRevX.6.041021}.  Interestingly, axial pseudo-fields $\boldsymbol{E} _{5}$ and $\boldsymbol{B} _{5}$ couple opposite chiral fermions with opposite signs, and hence the chirality can be tested by using conventional experimental probes such as electrical transport.  It is worth to mention that whereas electromagnetic potentials are gauge dependent and hence are not observables, the axial gauge potentials are quantum expectation values and thus produce gauge-invariant and observable effects. Indeed, an experimental realization of strain-induced pseudo-magnetic fields  was recently observed in strained crystals of Re-doped MoTe$_{2}$ \cite{PhysRevB.100.115105}. In this paper we aim to explore also nonlinear transport phenomena induced by pseudo-fields, in particular the longitudinal magnetoconductivity and the planar Hall effect. We clearly differentiate the contributions arising from the Berry curvature from those arising from the OMM of charge carriers.

The coupling of Weyl fermions with pseudo-gauge fields not only produces new interesting transport phenomena in Weyl semimetals, but also affects the well-known chiral anomaly. The inclusion of pseudo-fields in the semiclassical derivation of the chiral anomaly produces an interesting generalization of the anomaly equation, known in high-energy physics as the covariant anomaly \cite{PhysRevB.88.104412}. In the presence of genuine and pseudo-electromagnetic fields, the covariant anomaly equations reads
\begin{align}
\partial _{\mu} J ^{\mu} _{5} &= \frac{e^{2}}{2 \pi ^{2} \hbar ^{2}} ( \boldsymbol{E} \cdot \boldsymbol{B} + \boldsymbol{E} _{5} \cdot \boldsymbol{B} _{5} ) ,   \label{Chiral_Anomaly} \\  \partial _{\mu} J ^{\mu}  &= \frac{e^{2}}{2 \pi ^{2} \hbar ^{2}} ( \boldsymbol{E} \cdot \boldsymbol{B} _{5} + \boldsymbol{E} _{5} \cdot \boldsymbol{B} ) ,  \label{Charge_Anomaly}
\end{align}
where the axial current $J _{5} ^{\mu} $ measures the difference between currents of opposite chiralities and $J ^{\mu}$ is the total current. The breaking of charge conservation, as indicated by Eq. (\ref{Charge_Anomaly}),  implies that additional currents mus exist in the system to restore local charge conservation. In fact, in the case of WSMs, this problem is cured by including the conventional anomalous Hall current.  Nonlinear transport phenomena considered in this work also provides a testing ground for the chiral anomaly induced transport. In fact, we interpret our findings in terms of  the chiral anomaly with pseudo-fields.

This paper is organized as follows. In Sec. \ref{Kinetic} we use chiral kinetic theory to investigate the planar Hall effect in Weyl semimetals in the presence of pseudo-fields. We obtain a general expression for the magnetoconductivity tensor, separating in a clear fashion the contributions arising from the Berry curvature and the orbital magnetic moment. In Sec. \ref{PHE_Weyl} we evaluate such contributions for a simple linearly dispersing model of a WSM. We discuss the total and axial conductivities and interpret them in terms of the chiral anomaly with pseudo-fields. Section \ref{strain_section} is devoted to applications of our results to strained Weyl semimetals.  We conclude in Sec. \ref{conclusion}. All technical calculations are relegated to the Appendices.


\section{Kinetic theory approach} \label{Kinetic}

We will now investigate the PHE in Weyl semimetals by using the chiral kinetic theory, which is a topologically modified semiclassical Boltzmann formalism to describe the behavior of Weyl fermions for a finite chemical potential. Within this approach, the semiclassical equations of motion are extended to include  an anomalous velocity term arising from the Berry curvature, which acts as a magnetic field in reciprocal space \cite{RevModPhys.82.1959}. In the presence of electromagnetic fields (${\boldsymbol{E}}$ and ${\boldsymbol{B}}$) and axial pseudo-fields (${\boldsymbol{E}} _{5}$ and ${\boldsymbol{B}} _{5}$),  the semiclassical equations of motion for an electron wavepacket in a metal can be cast in the standard form \cite{Roy_2018}
\begin{align}
    \dot{{\boldsymbol{r}}} _{\alpha} &= \frac{1}{\hbar}  \nabla _{{\boldsymbol{k}}} \mathcal{E} _{\alpha} ({\boldsymbol{k}})  - \dot{{\boldsymbol{k}}} _{\alpha} \times {\boldsymbol{\Omega}} _{\alpha} ({\boldsymbol{k}}) , \label{r_dot_eq} \\[4pt] \hbar \dot{{\boldsymbol{k}}} _{\alpha} &= - e {\boldsymbol{E}} _{\chi} - e \dot{{\boldsymbol{r}}} _{\alpha} \times {\boldsymbol{B}} _{\chi} , \label{k_dot_eq}
\end{align}
where ${\boldsymbol{E}} _{\chi} = {\boldsymbol{E}} + \chi {\boldsymbol{E}} _{5}$ and ${\boldsymbol{B}} _{\chi} = {\boldsymbol{B}} + \chi {\boldsymbol{B}} _{5}$ are effective fields.  Elastic gauge fields $\boldsymbol{\mathcal{E}}$ and $\boldsymbol{\mathcal{B}}$, which couple to the Weyl fermions in a similar manner than genuine electromagnetic fields, are accounted by promoting  ${\boldsymbol{E}} \to {\boldsymbol{E}} + \boldsymbol{\mathcal{E}}$ and ${\boldsymbol{B}} \to {\boldsymbol{B}} + \boldsymbol{\mathcal{B}}$. The presence of $\chi$ in the definitions of the effective fields accounts for the fact that pseudo-fields couple opposite chiral fermions with opposite signs.  Here, $ {\boldsymbol{\Omega}} _{\alpha} ({\boldsymbol{k}}) = i \bra{\nabla _{{\boldsymbol{k}}} u _{\alpha} ({\boldsymbol{k}}) } \times \ket{ \nabla _{{\boldsymbol{k}}} u _{\alpha} ({\boldsymbol{k}}) }$ is the Berry curvature and $\mathcal{E} _{\alpha} ({\boldsymbol{k}}) = \mathcal{E} _{\alpha} ^{(0)} - {\boldsymbol{m}} _{\alpha} \cdot {\boldsymbol{B}} _{\chi} $ is the energy dispersion which includes a Zeeman-like correction due to the orbital magnetic moment ${\boldsymbol{m}} _{\alpha} ({\boldsymbol{k}}) = - i \frac{e}{2 \hbar } \bra{\nabla _{{\boldsymbol{k}}} u _{\alpha} ({\boldsymbol{k}}) } \times [ \hat{H} ({\boldsymbol{k}}) - \mathcal{E} _{\alpha} ^{(0)} ({\boldsymbol{k}}) ] \ket{ \nabla _{{\boldsymbol{k}}} u _{\alpha} ({\boldsymbol{k}}) } $  \cite{PhysRevB.53.7010, PhysRevB.59.14915}. Here, the Bloch states $\ket{u _{\alpha} ({\boldsymbol{k}})}$ are defined by $\hat{H}({\boldsymbol{k}}) \ket{u _{\alpha} ({\boldsymbol{k}})} = \mathcal{E} _{\alpha} ^{(0)} ({\boldsymbol{k}}) \ket{u _{\alpha} ({\boldsymbol{k}})} $ with $B _{\chi} = 0$. The subindex $\alpha$ stands collectively for the band index $s$ and the chirality index $\chi$.

As they are, the equations of motion (\ref{r_dot_eq}) and (\ref{k_dot_eq})  are reminiscent of the standard semiclassical equations by replacing the electromagnetic fields ${\boldsymbol{E}}$ and ${\boldsymbol{B}}$ by the pseudo-fields ${\boldsymbol{E}} _{\chi}$ and ${\boldsymbol{B}} _{\chi}$ solely; however, they do not immediately follow the semiclassical theory.  The wave-packet dynamics of electrons in crystals subject to perturbations varying slowly in space and time yields to generalized equations of motion that contain corrections which are accounted by a generalized Berry curvature defined in terms of two derivatives of the Bloch functions with respect to momentum (as the one defined above), position and time \cite{PhysRevB.53.7010, PhysRevB.59.14915}.  From the generalized equations of motion and assuming that the moment of the wavepacket is close to a Weyl node, a change of coordinate frame produces the equation of motion (\ref{r_dot_eq}) and (\ref{k_dot_eq}) \cite{Roy_2018}.

In the presence of impurity scattering the phenomenological transport equation can be written as \cite{Ashcroft76}
\begin{align}
    \left( \frac{\partial}{\partial t} + \dot{{\boldsymbol{r}}} _{\alpha} \cdot \nabla _{{\boldsymbol{r}}} + \dot{{\boldsymbol{k}}} _{\alpha} \cdot \nabla _{{\boldsymbol{k}}} \right) f _{\alpha} ({\boldsymbol{r}},{\boldsymbol{k}},t) = I _{\mbox{\scriptsize coll }} [f _{\alpha} ({\boldsymbol{r}},{\boldsymbol{k}},t)] , \label{Boltzmann_Eq}
\end{align}
where $f _{\alpha} ({\boldsymbol{r}},{\boldsymbol{k}},t)$ is the electron distribution function. The collision integral $I_{\text{coll}}$ accounts for the scattering mechanisms of the conduction electrons (such as impurity scattering effects, electron correlations, or scattering effects due to thermal vibrations of lattice ions). In the relaxation time approximation, the collision integral takes the simple form $I _{\mbox{\scriptsize{coll}}} [f _{\alpha}] = - \frac{f _{\alpha} - f _{\alpha} ^{\mbox{\tiny eq}}}{\tau ({\boldsymbol{k}}) } $, where $\tau ({\boldsymbol{k}})$ is the scattering time of quasiparticles and $f _{\alpha} ^{\mbox{\scriptsize eq}}$ is the equilibrium Fermi-Dirac distribution to be evaluated at the modified dispersion $\mathcal{E} _{\alpha} ({\boldsymbol{k}})$. Although the momentum dependence of $\tau$ covers a wide range of scattering processes, taking it as a constant parameter is still a good approximation that unveils interesting physics. In the following we take it as constant.

Here we are interested in stationary and homogeneous solutions to the Boltzmann equation (\ref{Boltzmann_Eq}). Using the equations of motion (\ref{r_dot_eq}) and (\ref{k_dot_eq}) we have
\begin{align}
- \tau \, \dot{{\boldsymbol{k}}} _{\alpha} \cdot \nabla _{{\boldsymbol{k}}}   f _{\alpha} ({\boldsymbol{k}} ) =  f _{\alpha} ({\boldsymbol{k}} ) - f _{\alpha} ^{\mbox{\scriptsize eq}} ({\boldsymbol{k}} ) . \label{Boltzmann_Eq2}
\end{align}
Next, we expand the distribution function in powers of the electromagnetic fields. Keeping only linear order dependence on the electric field, the nonequilibrium distribution function becomes
\begin{align}
f _{\alpha} = f _{\alpha} ^{\mbox{\scriptsize eq}}  + e \tau D _{\alpha }  \left[   {\boldsymbol{v}} _{\alpha} \cdot {\boldsymbol{E}} _{\chi} + \frac{e}{\hbar} ({\boldsymbol{E}} _{\chi} \cdot {\boldsymbol{B}} _{\chi} ) ( {\boldsymbol{v}} _{\alpha} \cdot {\boldsymbol{\Omega}} _{\alpha} ) \right]  \frac{\partial f _{\alpha} ^{\mbox{\scriptsize eq}}  }{ \partial \mathcal{E} _{\alpha}  } , \label{Boltzmann_Eq_Sol}
\end{align}
where ${\boldsymbol{v}} _{\alpha} ({\boldsymbol{k}} ) = \tfrac{1}{\hbar}  \nabla _{{\boldsymbol{k}}} \mathcal{E} _{\alpha} ({\boldsymbol{k}} ) $ is the band velocity of Bloch electrons and $D _{\alpha } ({\boldsymbol{k}} ) = [1 + \tfrac{e}{\hbar} ({\boldsymbol{B}} _{\chi} \cdot {\boldsymbol{\Omega}} _{\alpha} ) ]^{-1} $ is the modification factor of the phase space volume element \cite{PhysRevLett.97.026603, doi:10.1142/S0217984906010573}. In these expressions we have omitted all momentum dependencies for simplicity.

In the absence of any thermal and chemical potential gradients, the charge density current  for a single $\alpha$ (i.e. band, chirality, etc.) can be written as ${\boldsymbol{J}} _{\alpha} = - e \int \frac{d ^{3} {\boldsymbol{k}}}{(2 \pi ) ^{3} } D _{\alpha } ^{-1} \dot{{\boldsymbol{r}}} _{\alpha} ({\boldsymbol{k}} ) f _{\alpha} ({\boldsymbol{k}} ) $, accounting for the modified density of states due to the phase space factor $D _{\alpha }$. Substituting the nonequilibrium distribution function (\ref{Boltzmann_Eq_Sol}) into this equation and keeping the linear term in the electric field, but neglecting the chiral magnetic and anomalous Hall effect contributions, we now arrive at the expression for the magnetoconductivity tensor for a single $\alpha$:
\begin{align}
\sigma ^{(\alpha)} _{ij} ({\boldsymbol{B}} _{\chi})  &= - e ^{2} \tau \int \frac{d ^{3} {\boldsymbol{k}}}{(2 \pi ) ^{3} } D _{\alpha} \left[  v _{\alpha i } + \frac{e}{\hbar}  ( {\boldsymbol{v}} _{\alpha} \cdot {\boldsymbol{\Omega}} _{\alpha} ) B _{\chi i} \right] \notag \\ &  \phantom{===} \times \left[  v _{\alpha j } + \frac{e}{\hbar}  ( {\boldsymbol{v}} _{\alpha} \cdot {\boldsymbol{\Omega}} _{\alpha} ) B _{\chi j} \right] \frac{\partial f _{\alpha} ^{\mbox{\scriptsize eq}} (\mathcal{E} _{\alpha}) }{ \partial \mathcal{E} _{\alpha}  } , \label{Conductivity_Tensor}
\end{align}
which includes the effects of the Berry curvature and the orbital magnetic moment. Our main goal in this paper is to evince that the orbital magnetic moment contributes in most the same fashion that the Berry curvature to the PHE. To distinguish these contributions we write the conductivity tensor (\ref{Conductivity_Tensor}), in the weak-field limit, as the sum of three terms:
\begin{align}
\sigma ^{(\alpha)} _{ij} ({\boldsymbol{B}} _{\chi})  &= \sigma ^{(0, \alpha)} _{ij} + \sigma ^{( \Omega , \alpha)} _{ij} ({\boldsymbol{B}} _{\chi}) +  \sigma ^{( m , \alpha)} _{ij} ({\boldsymbol{B}} _{\chi}) , \label{totalsigma}
\end{align}
where the first term, 
\begin{align}
    \sigma ^{(0, \alpha)} _{ij} = - e ^{2} \tau \int \frac{d ^{3} {\boldsymbol{k}}}{(2 \pi ) ^{3} }  v _{\alpha i } ^{(0)} v _{\alpha j} ^{(0)} \,  \frac{\partial f _{\alpha} ^{\mbox{\scriptsize eq}} (\mathcal{E} _{\alpha} ^{(0)} ) }{ \partial \mathcal{E} _{\alpha} ^{(0)} } ,  \label{Conductivity_Tensor_0B}
\end{align}
is the conductivity in the absence of the magnetic field (i.e. for $\boldsymbol{B} =\boldsymbol{B} _{5} = {\bf{0}}$) and the second term,
\begin{align}
\sigma ^{( \Omega , \alpha)} _{ij} ({\boldsymbol{B}} _{\chi})  = - \frac{e ^{4} \tau}{\hbar ^{2}} \int \frac{d ^{3} {\boldsymbol{k}}}{(2 \pi ) ^{3} } Q _{\alpha i } \, Q _{\alpha j } \,  \frac{\partial f _{\alpha} ^{\mbox{\scriptsize eq}} (\mathcal{E} _{\alpha} ^{(0)}  ) }{ \partial \mathcal{E} _{\alpha} ^{(0)}  } ,  \label{Conductivity_Tensor_Berry}
\end{align}
with ${\boldsymbol{Q}} _{\alpha}  = {\boldsymbol{\Omega}} _{\alpha}  \times ( \boldsymbol{v} _{\alpha} ^{(0)}   \times {\boldsymbol{B}} _{\chi}  )$, is the contribution arising from the Berry curvature solely. In Eqs. (\ref{Conductivity_Tensor_0B}) and (\ref{Conductivity_Tensor_Berry}), $\mathcal{E} _{\alpha} ^{(0)} ({\boldsymbol{k}} )$ is the band energy without the Zeeman-like correction and ${\boldsymbol{v}} _{\alpha} ^{(0)} ({\boldsymbol{k}} ) = \tfrac{1}{\hbar} \nabla _{{\boldsymbol{k}}} \mathcal{E} _{\alpha} ^{(0)} ({\boldsymbol{k}} ) $ is the corresponding band velocity.  In deriving Eq. (\ref{Conductivity_Tensor_Berry}) we have expanded the phase space volume factor $D _{\alpha } ({\boldsymbol{k}} )$ up to second order in the magnetic field.

The contribution from the orbital magnetic moment, $\sigma ^{( m , \alpha)} _{ij} ({\boldsymbol{B}} _{\chi})$ , appears in various ways. In the presence of a magnetic field, on the one hand,  the energy dispersion is corrected by $\mathcal{E} _{\alpha} ^{(m)} ({\boldsymbol{k}} ) = -  {\boldsymbol{m}} _{\alpha} ({\boldsymbol{k}} ) \cdot {\boldsymbol{B}} _{\chi}$ and consequently the band velocity becomes corrected  as ${\boldsymbol{v}} _{\alpha} ^{(m)} ({\boldsymbol{k}} ) =  \tfrac{1}{\hbar} \nabla _{{\boldsymbol{k}}} \mathcal{E} _{\alpha} ^{(m)} ({\boldsymbol{k}} )$. On the other hand, for a weak magnetic field, the equilibrium distribution function $f _{\alpha} ^{\mbox{\scriptsize eq}} (\mathcal{E} _{\alpha} )$ and the phase space volume factor $D _{\alpha} ({\boldsymbol{k}} ) $ can be Taylor expanded up to the second power of the magnetic field (see Appendix \ref{App_OMM}). Taking into account these terms in Eq. (\ref{Conductivity_Tensor}) and substracting the contributions $\sigma ^{(0, \alpha)} _{ij}$ and $\sigma ^{( \Omega , \alpha)} _{ij} ({\boldsymbol{B}} _{\chi})$, one gets

\begin{widetext}
\begin{align}
\sigma _{ij} ^{(\chi , m )} ({\boldsymbol{B}} _{\chi}) =  \frac{2 e ^{3} \tau}{\hbar} \! \int \! \frac{d ^{3} {\boldsymbol{k}} }{(2 \pi ) ^{3}} \,  \bigg[    {Q} _{\alpha i} v _{\alpha j} ^{(m)}   + \frac{1}{e} \mathcal{E} _{\alpha} ^{(m)}  \nabla _{{\boldsymbol{k}}} \cdot  \boldsymbol{T} _{\alpha ij} + \frac{1}{2} \mathcal{E} _{\alpha} ^{(m)}   \boldsymbol{B} _{\chi} \cdot \boldsymbol{V} _{\alpha ij }  \frac{\partial  }{ \partial \mathcal{E} _{\alpha} ^{(0)}  } \bigg] \frac{\partial f _{\alpha} ^{\mbox{\scriptsize eq}} (\mathcal{E} _{\alpha} ^{(0)}  ) }{ \partial \mathcal{E} _{\alpha} ^{(0)}  }   , \label{Conductivity_Tensor_OMM}
\end{align}
where we have defined the tensors
\begin{align}
\boldsymbol{T} _{\alpha ij} = \frac{e}{\hbar}  \boldsymbol{\Omega} _{\alpha } B _{\chi i}   v _{\alpha j} ^{(0)} + \frac{1}{2}   \hat{\boldsymbol{e}} _{i} v _{\alpha j } ^{(m)} , \qquad  \boldsymbol{V} _{\alpha ij} = \boldsymbol{\Omega} _{\alpha} v _{\alpha i} ^{(0)} v _{\alpha j} ^{(0)} - \frac{1}{2e} \boldsymbol{m} _{\alpha}  \partial _{k _i} v _{\alpha j} ^{(0)}  . \label{Tensors}
\end{align}
A detailed derivation of the formula (\ref{Conductivity_Tensor_OMM}) is presented in the Appendix \ref{App_OMM}. 

\end{widetext}

\section{Planar Hall effect with pseudofields in Weyl semimetals} \label{PHE_Weyl}

In this section we investigate the PHE in Weyl semimetals by using the semiclassical approach developed in the previous section. To this end we consider a simple model of a WSM consisting of two Weyl nodes of opposite chiralities separated in momentum and energy, ignoring the nonuniversal corrections due to band bending far away from the nodes. The low energy Hamiltonian for each Weyl node can be expressed as 
\begin{align}
\hat{H} _{\chi} ({\boldsymbol{k}}) = \chi \hbar v _{F} \boldsymbol{\sigma} \cdot {\boldsymbol{k}} + b _{0 \chi} ,  \label{Hamiltonian}
\end{align}
where $v _{F}$ is the Fermi velocity, $\chi = \pm 1$ specifies the chirality,  $\boldsymbol{\sigma}$ is the vector of the Pauli matrices, ${\boldsymbol{k}}$ is the momentum measured relative to the Weyl point and $b _{0 \chi}$ denotes the energy shift of the node with chirality $\chi$.

\begin{figure}
    \centering
    \includegraphics[width=0.4\textwidth]{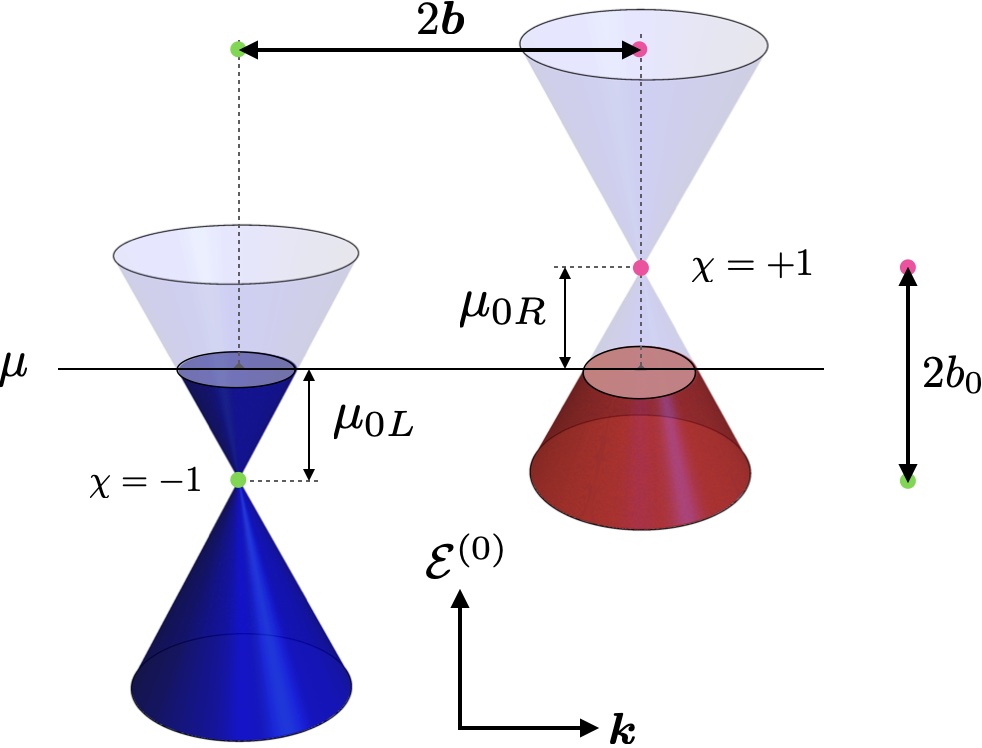}
    \caption{Low-energy spectrum of a Weyl semimetal with two bulk Weyl nodes of different chiralities separated in momentum space by $2 {\boldsymbol{b}}$ and shifted in energy by $2b _{0}$.} \label{Fig_Cones}
\end{figure}

The corresponding energy dispersion is $\mathcal{E} ^{(0)} _{\alpha} ({\boldsymbol{k}}) = b _{0 \chi} + s \hbar v _{F} k$, where $s = \pm 1 $ is the band index. As a result, the band velocity becomes ${\boldsymbol{v}} _{s} ^{(0)} ({\boldsymbol{k}}) = s v _{F} \hat{{\boldsymbol{k}}}$, where $\hat{{\boldsymbol{k}}}$ is the unit vector along ${\boldsymbol{k}}$. Using the Bloch states, it is straightforward to obtain the Berry curvature and the orbital magnetic moment:
\begin{align}
{\boldsymbol{\Omega}} _{\alpha} ({\boldsymbol{k}}) = - s \chi \frac{\hat{{\boldsymbol{k}}}}{2k ^{2}} , \qquad {\boldsymbol{m}} _{\chi} ({\boldsymbol{k}}) = - \chi e v_{F} \frac{\hat{{\boldsymbol{k}}}}{2k} , \label{B_Curvature}
\end{align}
respectively. In the presence of an effective magnetic magnetic field ${\boldsymbol{B}} _{\chi}$, the energy dispersion is corrected by $ \mathcal{E} _{\chi} ^{(m)} ({\boldsymbol{k}})= \frac{\chi e v_{F}}{2 k} \hat{{\boldsymbol{k}}} \cdot {\boldsymbol{B}} _{\chi}$, which implies a correction to the band velocity as
\begin{align}
{\boldsymbol{v}} _{\chi } ^{(m)} ({\boldsymbol{k}}) =  \frac{ \chi e v _{F}  }{2 \hbar   }  \frac{ {\boldsymbol{B}} _{\chi} - 2  \hat{{\boldsymbol{k}}} ( \hat{{\boldsymbol{k}}} \cdot {\boldsymbol{B}} _{\chi} )}{k ^{2}}   .   \label{OMM_velocity_main}
\end{align}
For a given node of chirality $\chi$, the band index is determined by the sign of the
difference between the chemical potential and the energy shift of the node, i.e. $s = \mbox{sgn} (\mu - b _{0 \chi})$.  This is so since $\mu > b _{0 \chi}$ ($\mu < b _{0 \chi}$) implies $s = 1$ ($s=-1$), as depicted in Fig.  \ref{Fig_Cones}. So, we define $\mu _{\chi} = s \mu _{0 \chi}$, with $\mu _{0 \chi} = \vert \mu - b _{0 \chi} \vert > 0$. {Here we work at finite temperature $T$.}

With the above information we are able to apply the semiclassical formulas (\ref{Conductivity_Tensor_0B})-(\ref{Conductivity_Tensor_OMM}) to compute the different contributions to the planar Hall effect in Weyl semimetals. Details of technical computations are relegated to the Appendix \ref{Det_Calc} and here we present only the final results.  Interestingly, all of them become independent of the band index $s$, so we make the replacement $\alpha \to \chi$ in the following expressions.

The $\boldsymbol{B} _{\chi}$-independent conductivity (\ref{Conductivity_Tensor_0B}) takes the simple form
\begin{align}
\sigma ^{(0, \chi)} _{ij} (T)= \frac{e ^{2} \mu _{0 \chi } ^{2} \tau }{6 \pi ^{2} \hbar ^{3} v _{F} } \delta_{ij}  \,  f _{2} (\Lambda _{\chi} )  , \label{B0_Conductivity}
\end{align}
with $\Lambda _{\chi} \equiv k _{B} T / \mu _{0 \chi }$ and where we have introduced the function
\begin{align}
{ f _{n} (y) \equiv  \frac{1}{y} \int _{0} ^{\infty} dx \, x ^{n} \,   \frac{e ^{ (x - 1 )/ y }}{ \left[ 1 + e ^{ (x - 1 )/ y  } \right] ^{2} } } .  \label{F_function}
\end{align}
Interestingly,  the conductivity (\ref{B0_Conductivity}) vanishes at the neutrality point ($\mu _{0 \chi } = 0$). Furthermore, at zero temperature one can further verify that $\lim _{T \to 0} f _{n} (\Lambda _{\chi} ) = 1$.

The Berry curvature contribution (\ref{Conductivity_Tensor_Berry}) becomes
\begin{align}
\sigma ^{( \Omega , \chi )} _{ij} ( \boldsymbol{B} _{\chi},T )  &= \! \frac{e ^{4} v _{F} ^{3} \tau}{120 \pi ^{2} \hbar \mu _{0 \chi} ^{2} } \left( \delta _{ij} B _{\chi} ^{2}  + 7 B _{\chi i} B _{\chi j} \right) f _{-2} (\Lambda _{\chi} ) ,   \label{Conductivity_Tensor_Berry_Final}
\end{align}
while the result from orbital magnetic part is
\begin{align}
\sigma _{ij} ^{(m , \chi )} ( \boldsymbol{B} _{\chi} ,T) = \!  \frac{- e ^{4} v _{F} ^{3} \tau}{120 \pi ^{2} \hbar \mu _{0 \chi} ^{2} } \left( 3 \delta _{ij} B _{\chi} ^{2} +  B _{\chi i} B _{\chi j} \right) f _{-2} (\Lambda _{\chi} ) , \label{Conductivity_Tensor_MagMoment_Final}
\end{align}
where $B _{\chi} ^{2} = \boldsymbol{B} _{\chi} \cdot \boldsymbol{B} _{\chi}$. The conductivities of the opposite chiralities are related by symmetry properties dictated by the definition of the effective magnetic field $\boldsymbol{B} _{\chi}$, namely, $\sigma ^{( \xi , \chi )} _{ij} ( \boldsymbol{B} , \boldsymbol{B} _{5} ,T) = \sigma ^{( \xi , - \chi )} _{ij} ( - \boldsymbol{B} ,   \boldsymbol{B} _{5} ,T)$ and $\sigma ^{( \xi , \chi )} _{ij} ( \boldsymbol{B} , \boldsymbol{B} _{5} ,T) = \sigma ^{( \xi , - \chi )} _{ij} (  \boldsymbol{B} , - \boldsymbol{B} _{5},T )  $, where $\xi = \Omega , m$. If either $\boldsymbol{B} $ or $ \boldsymbol{B} _{5}$ vanishes, the contribution from both chiralities are the same. Besides, we observe that the transverse conductivities does not satisfy the usual antisymmetry relation ($\sigma _{xy} = - \sigma _{yx}$) displayed by Hall effect systems, since in this case the transverse conductivity does not stem from Lorentz force, but from the chiral anomaly.

\begin{figure}
    \centering
    \includegraphics[width=0.38\textwidth]{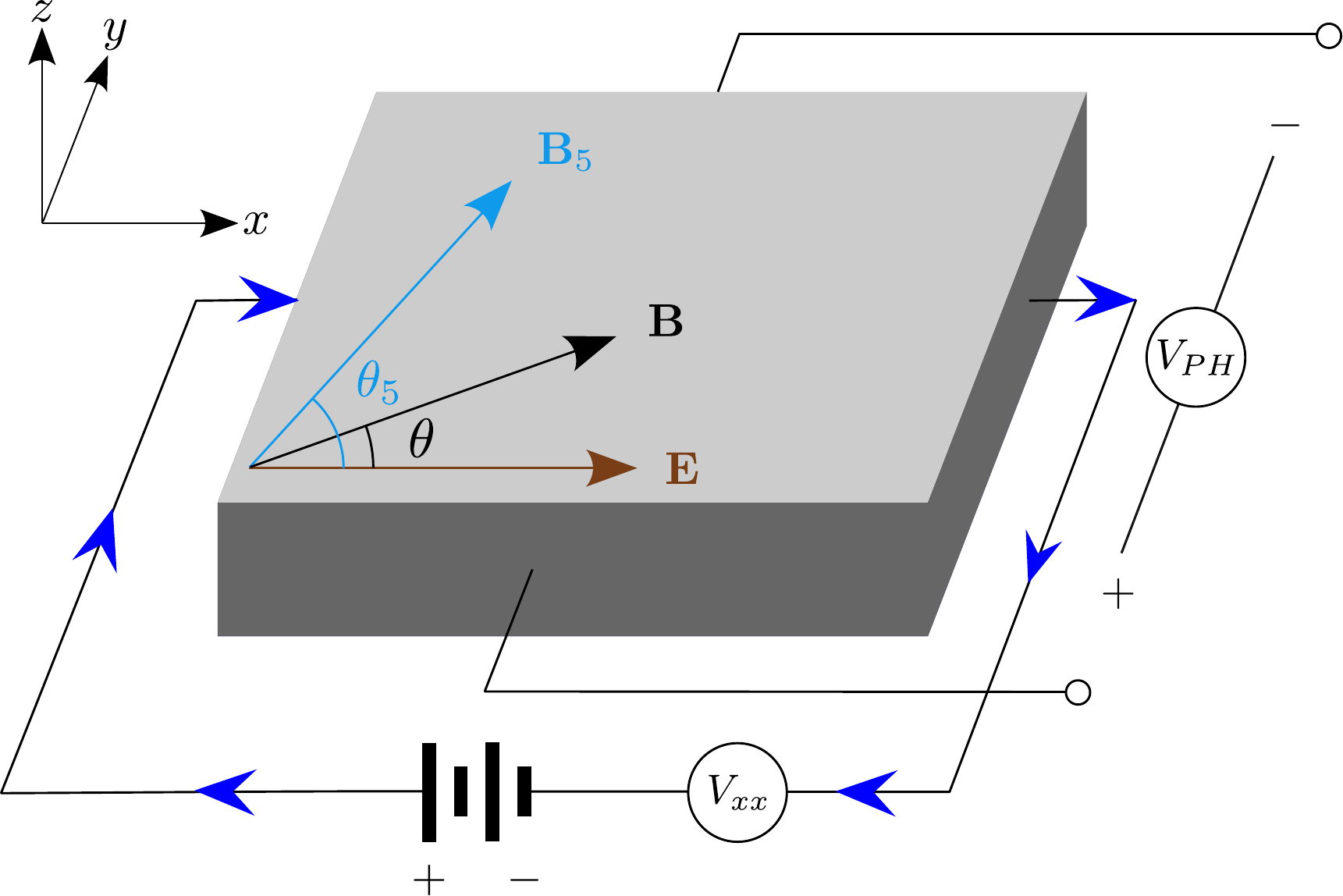}
    \caption{Prototypical experimental setup employed to measure the longitudinal and planar Hall voltages generated by in-plane electromagnetic fields and   pseudo fields.} \label{Setup}
\end{figure}

The Berry curvature induced conductivity (\ref{Conductivity_Tensor_Berry_Final}) has been discussed recently in a variety of papers and it has been regarded as a direct consequence of the chiral anomaly \cite{PhysRevLett.119.176804, PhysRevB.100.115139, PhysRevB.102.121105, PhysRevB.99.115121, PhysRevB.102.205107}. Our formula (\ref{Conductivity_Tensor_Berry_Final}) generalizes the previously reported results, where only some components were computed. Besides, here we report a general formula (\ref{Conductivity_Tensor_MagMoment_Final}) for the orbital magnetic moment induced conductivity, which as far as we know, it has not been reported yet. In order to elucidate the importance of the orbital magnetic moment contribution, we next explore the angular dependence of both the longitudinal magnetoconductivity and the planar Hall conductivity for the different chiralities. To this end,  we introduce the normalized conductivity tensor for the Weyl node with chirality $\chi$:
\begin{align}
\Sigma _{ij} ^{( \chi )}  (\boldsymbol{B} _{\chi},T) = \frac{ \sigma ^{( \chi )} _{ij} (\boldsymbol{B} _{\chi},T) - \sigma _{0} ^{(\chi)} (T) \delta _{ij} }{ \sigma _{0} ^{(\chi)} (T) } , \label{Sigma_tensor}
\end{align}
where $\sigma ^{( \chi )} _{ij} (\boldsymbol{B} _{\chi},T)$ is the total conductivity given by Eq.  (\ref{totalsigma}) and $\sigma _{0} ^{(\chi)} (T) \equiv \frac{e ^{2} \mu _{0 \chi } ^{2} \tau }{6 \pi ^{2} \hbar ^{3} v _{F} }  \,  f _{2} (\Lambda _{\chi} )  $ is the longitudinal conductivity. Note that $\Sigma _{ij} ^{( \chi )}$ isolates the joint contributions from the Berry curvature and the orbital magnetic moment of charge carriers. The tensor (\ref{Sigma_tensor}) inherits the symmetries of $\sigma ^{( \Omega , \chi )} _{ij}$ and $\sigma ^{( m, \chi )} _{ij}$. Now we assume an electric field pointing along the $x$-axis, i.e. $\boldsymbol{E} = E \hat{\boldsymbol{e}} _{x}$, and restrict the magnetic field to be in the $xy$-plane, i.e. $\boldsymbol{B} = B ( \cos \theta \hat{\boldsymbol{e}} _{x} + \sin \theta \hat{\boldsymbol{e}} _{y})$. We also assume that the applied strain induces a pseudomagnetic field lying in the $xy$-plane, i.e. $\boldsymbol{B} _{5} = B _{5} ( \cos \theta _{5} \hat{\boldsymbol{e}} _{x} + \sin \theta _{5} \hat{\boldsymbol{e}} _{y})$ and vanishing axial electric field $\boldsymbol{E} _{5} = \boldsymbol{0}$, as depicted in Fig. \ref{Setup}. Later we will discuss the impact of a nonzero $\boldsymbol{E} _{5}$ upon the longitudinal and planar Hall currents.

\begin{figure}
    \centering
	\includegraphics[width=0.22\textwidth]{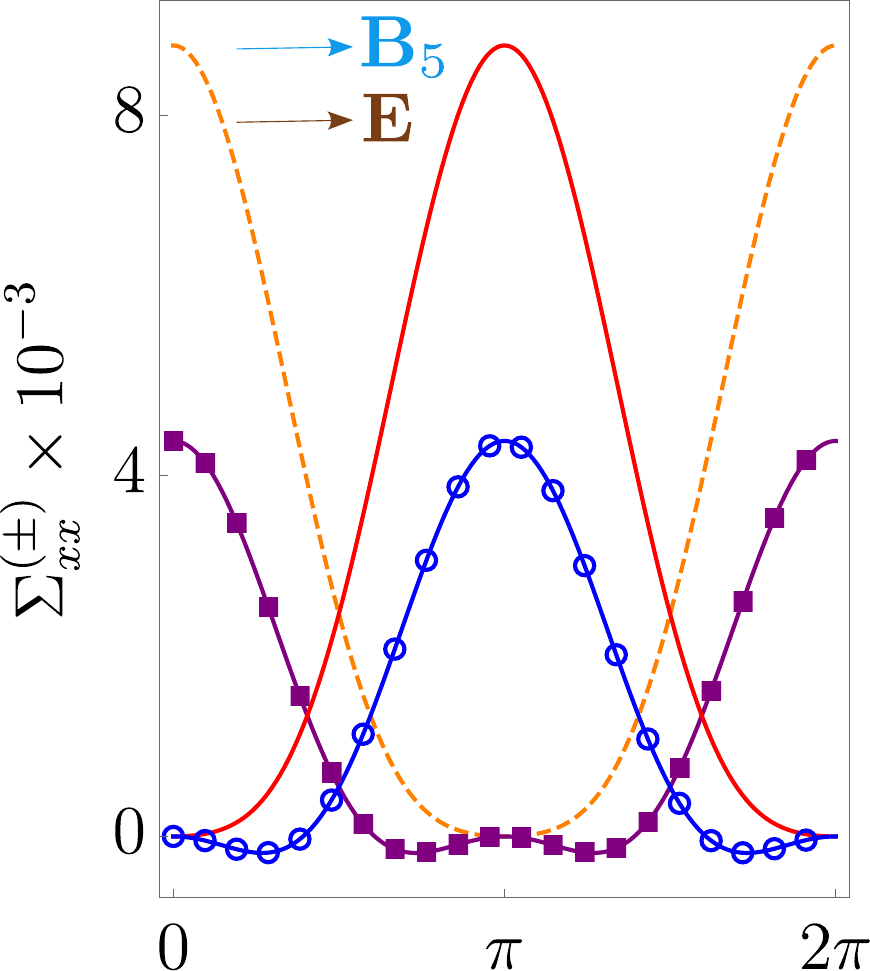} 
	\includegraphics[width=0.23\textwidth]{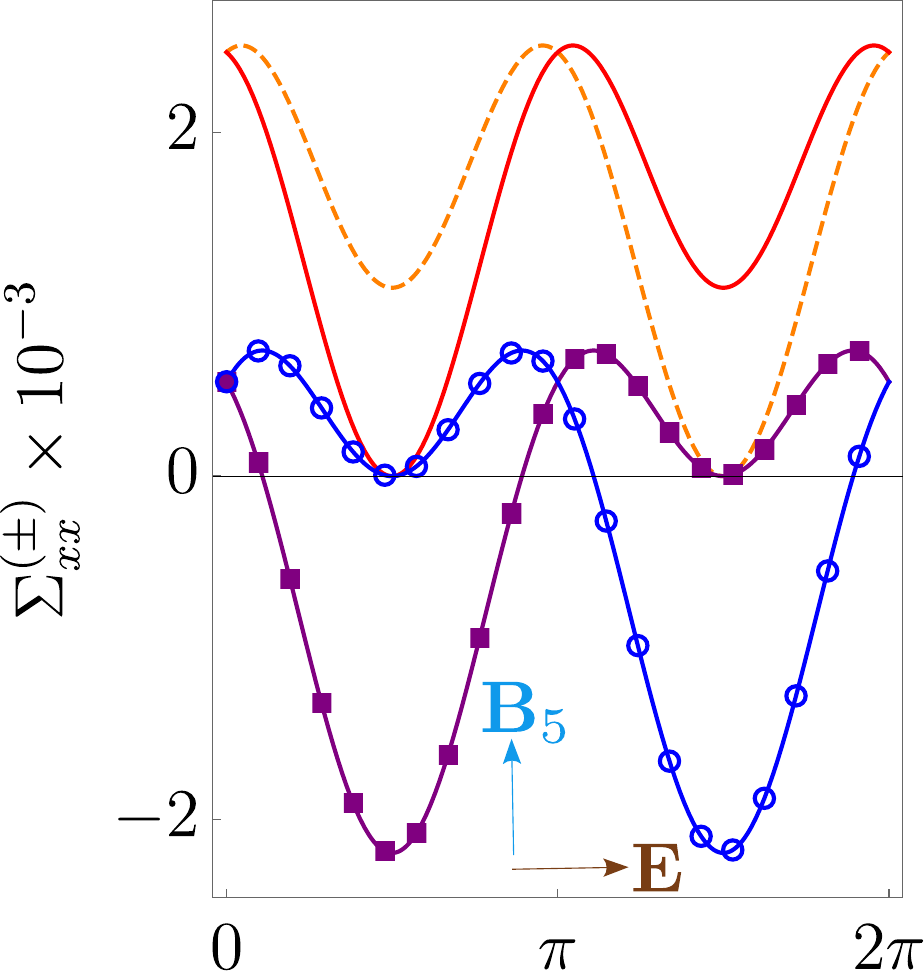}
	\\[6pt] \includegraphics[width=0.23\textwidth]{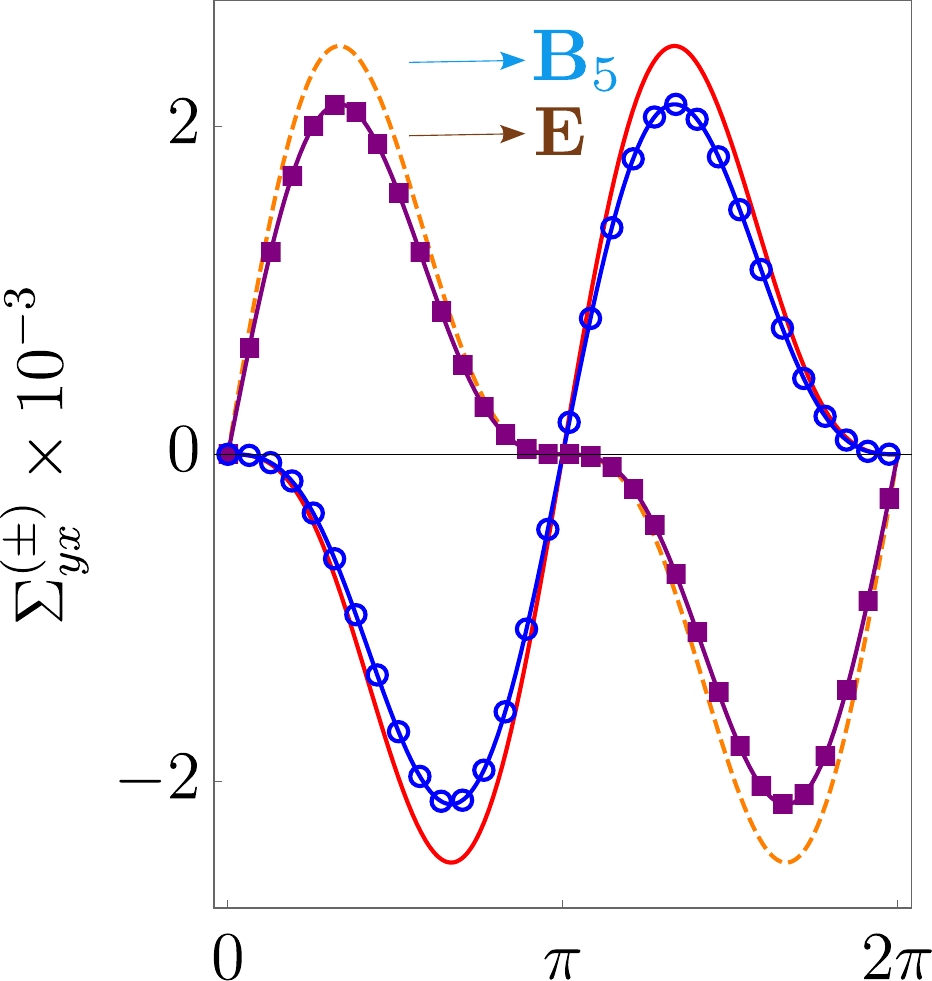}
	\includegraphics[width=0.23\textwidth]{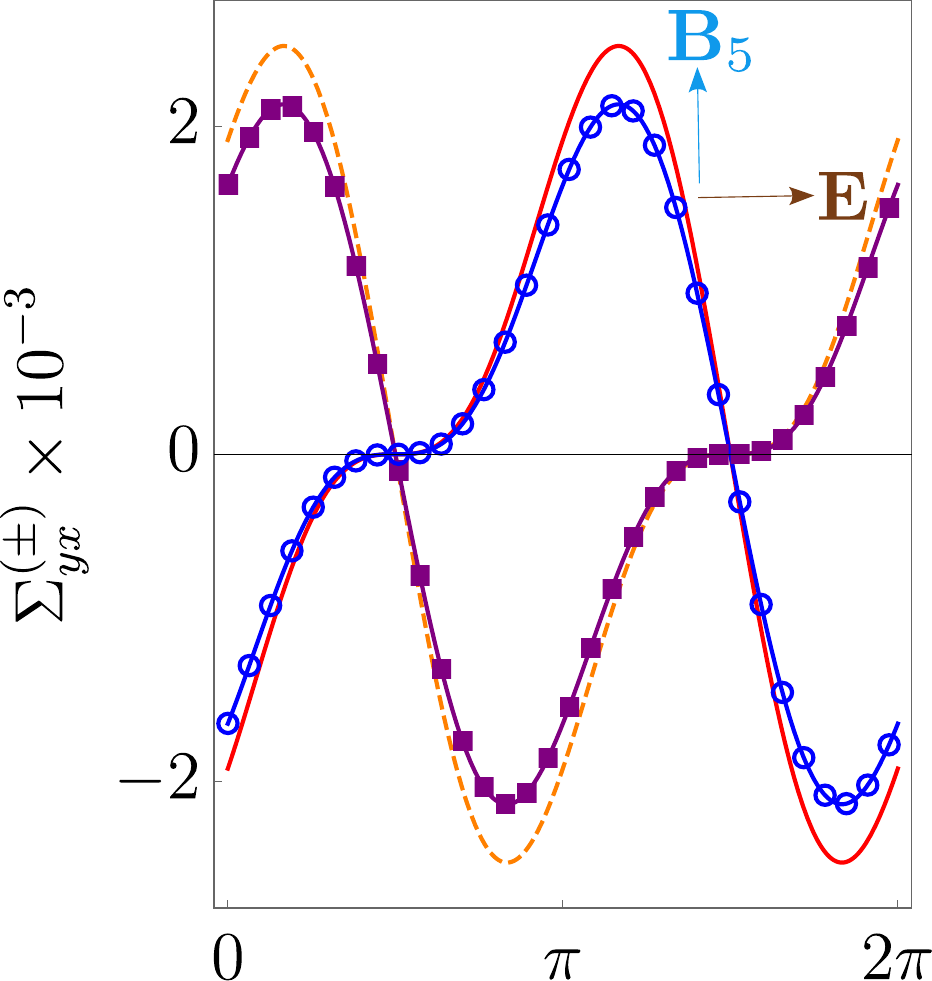}
     \caption{Angular dependence of the longitudinal (upper panel) and planar Hall (lower panel) conductivities for $\theta _{5} = 0$ (left panel) and $\theta _{5} = \pi /2$ (right panel). The  dashed orange (continuous red) line shows the Berry curvature contribution (normalized by the longitudinal conductivity) for the chirality $\chi = +1$ ($\chi = -1$), while the purple (blue) line with square (circle) markers shows the full conductivities including the orbital magnetic moment contribution.   } \label{Conduc_Chiralities}
\end{figure}

To discuss the longitudinal and planar Hall responses in a realistic WSM, it is convenient to consider the precise values of the parameters appearing in our expressions and verify first the validity of the chiral kinetic theory. We take $B=B_{5}=0.5$T and use  typical parameters for a Weyl semimetal such as TaAs:  $v_{F} = 3 \times 10 ^{5}$m/s, $b _{0 \chi} = 0$, $\mu=20$meV \cite{PhysRevX.5.011029, PhysRevX.5.031023}, and $\tau \sim 10 ^{-13}$s \cite{Xiong_2016, 10.1038/nphys3372}.  Therefore, the Boltzmann formalism is valid because $\omega _{\mbox{\scriptsize C}} \tau \sim 0.08 \ll 1$, where $\omega _{\mbox{\scriptsize C}} = eB/m^{\ast} c$ is the cyclotron frequency and we have used $m^{\ast} \sim 0.11 m _{e}$ \cite{PhysRevX.5.011029, PhysRevX.5.031023} and $B \sim 0.5$T. One can further verify that near the nodes, the corrections induced by the orbital magnetic moment to the energy satisfies $\mathcal{E} _{\chi} ^{(m)} \ll \mathcal{E} _{\alpha} ^{(0)}$ and $(e / \hbar) B \Omega \ll 1$, thus validating the expansions performed in Section \ref{Kinetic}.  {Here we take $T=2$K, which corresponds to the temperature at which the angular dependence of the planar Hall conductivity was observed in topological insulators \cite{10.1038/s41467-017-01474-8}.  This value, together with $\mu=20$meV for TaAs, implies that $\Lambda _{\chi} = 0.0086$ and hence $f _{-2} (\Lambda _{\chi} )= 1.00073 \approx 1$. Therefore, in the following we safely take $f _{-2} (\Lambda _{\chi} )= 1$ for definiteness, which is appropriate at low temperatures.} In Fig. \ref{Conduc_Chiralities} we plot the normalized longitudinal (upper panel) and planar Hall (lower panel) conductivities, given by Eq. (\ref{Sigma_tensor}), as a function of the angle $\theta$ and fixed values $\theta _{5} = 0$ (left panel) and $\theta _{5} = \pi/2$ (right panel).  In  these plots, the  dashed orange (continuous red) line shows the Berry curvature contribution (\ref{Conductivity_Tensor_Berry_Final}) normalized by the longitudinal conductivity $\sigma _{0} ^{(\chi)}$, i.e. $\sigma ^{( \Omega , \chi )} _{ij} / \sigma _{0} ^{(\chi)}$, for the chirality $\chi = + 1 $ ($\chi = - 1 $).  The purple (blue) line with square (circle) markers shows the normalized conductivity $\Sigma _{ij} ^{( \chi )} $ as defined in Eq.  (\ref{Sigma_tensor}) for the chirality $\chi = + 1 $ ($\chi = - 1 $). These plots evince the importance of the orbital magnetic moment upon the longitudinal and planar Hall conductivities. In fact, as we can see in Fig. \ref{Conduc_Chiralities}, the longitudinal conductivity is more sensitive to the OMM than the planar Hall conductivity. However, this fact does not mean that such contribution is zero in the planar Hall conductivity. Clearly, for $\theta _{5}=0$, the OMM contribution is negligible near to $\theta = 0 ,\pi $, but becomes important near to some specific angles, namely, at $\theta ^{\ast} = \pi /3, \, 5 \pi /3$ for $\chi = +1$ and  at $\theta ^{\ast} = 2 \pi /3, 4 \pi /3$ for $\chi = -1$. For $\theta _{5} = \pi /2$, however, the OMM contribution approaches zero near to $\theta = \pi /2, \, 3 \pi /2 $, and becomes appreciable around $\theta ^{\ast} = 7\pi /6, \, 11 \pi /6$ for $\chi = +1$ and  around $\theta ^{\ast} = \pi /6, \, 5 \pi /6$ for $\chi = -1$. The position of these critical angles $\theta ^{\ast}$ varies according to the ratio $B _{5}/B$ as well as the direction of $\boldsymbol{B} _{5}$. In fact, it becomes determined by the equation:
\begin{align}
\cos (2 \theta ^{\ast}) + \chi (B _{5} / B) \cos (\theta ^{\ast} + \theta _{5} ) = 0 .
\end{align}

Following the definition of the current for fermions with chirality $\chi$ in the presence of effective electromagnetic fields, $J _{ i}  ^{(\chi )} = \sigma ^{(\chi )} _{ij} E _{\chi j}$, with the conductivity given by Eq. (\ref{Conductivity_Tensor}), we now define the total and axial currents by $\boldsymbol{J} = \sum _{\chi = \pm 1} \boldsymbol{J} ^{(\chi )} $ and $\boldsymbol{J} _{5} = \sum _{\chi = \pm 1} \chi \boldsymbol{J} ^{(\chi )} $, respectively. These suggest the definition of the total and the chiral conductivities as follows:
\begin{align}
\sigma _{ij} = \sum _{\chi = \pm 1} \sigma ^{(\chi)} _{ij}  , \qquad \sigma  _{5ij}  = \sum _{\chi = \pm 1} \chi \sigma ^{(\chi)} _{ij}  ,  \label{Chiral_conductivities}
\end{align}
each in order, such that the total and axial currents take the simple form
\begin{align}
J _{i} = \sigma _{ij} ({\boldsymbol{B}} , {\boldsymbol{B}} _{5} ) \,  E _{j} + \sigma  _{5ij} ({\boldsymbol{B}} , {\boldsymbol{B}} _{5} ) \, E _{5j} ,  \label{Total_Current} \\[5pt] J _{5i} = \sigma _{5ij} ({\boldsymbol{B}} , {\boldsymbol{B}} _{5} ) \, E _{j} + \sigma  _{ij} ({\boldsymbol{B}} , {\boldsymbol{B}} _{5} ) \, E _{5j} , \label{Axial_Current}
\end{align}
respectively.

\begin{figure}
    \centering
    \includegraphics[width=0.23\textwidth]{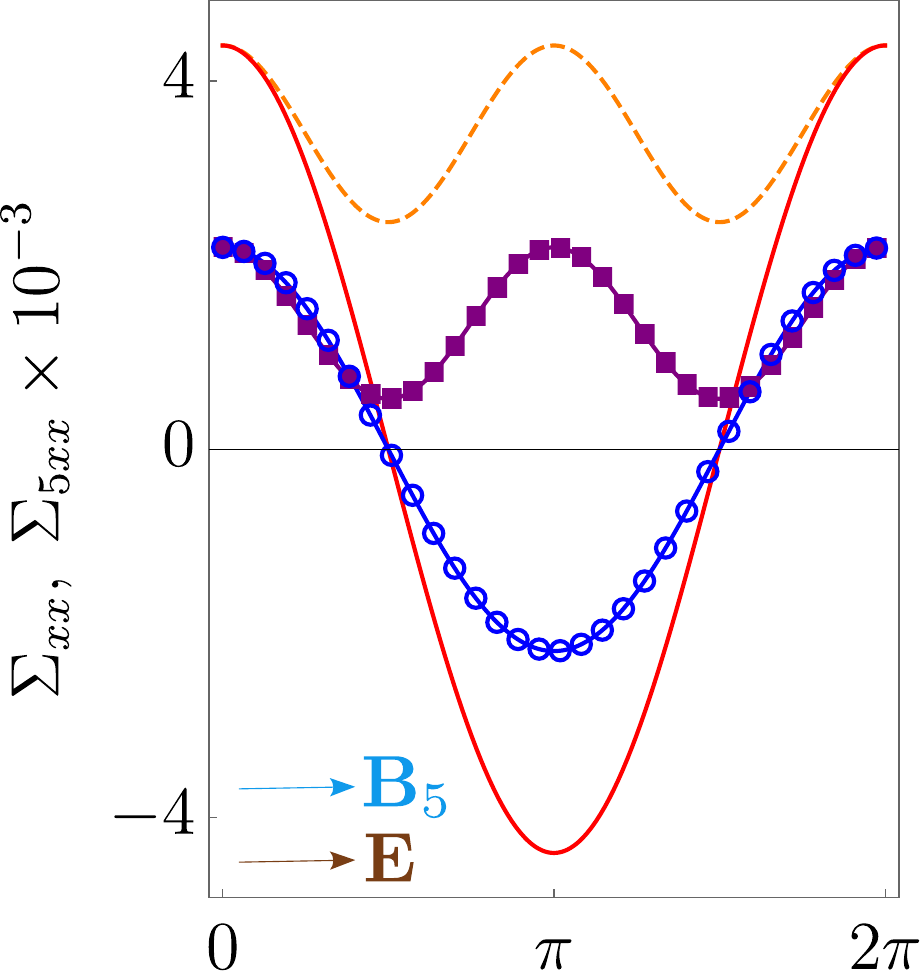}
     \includegraphics[width=0.23\textwidth]{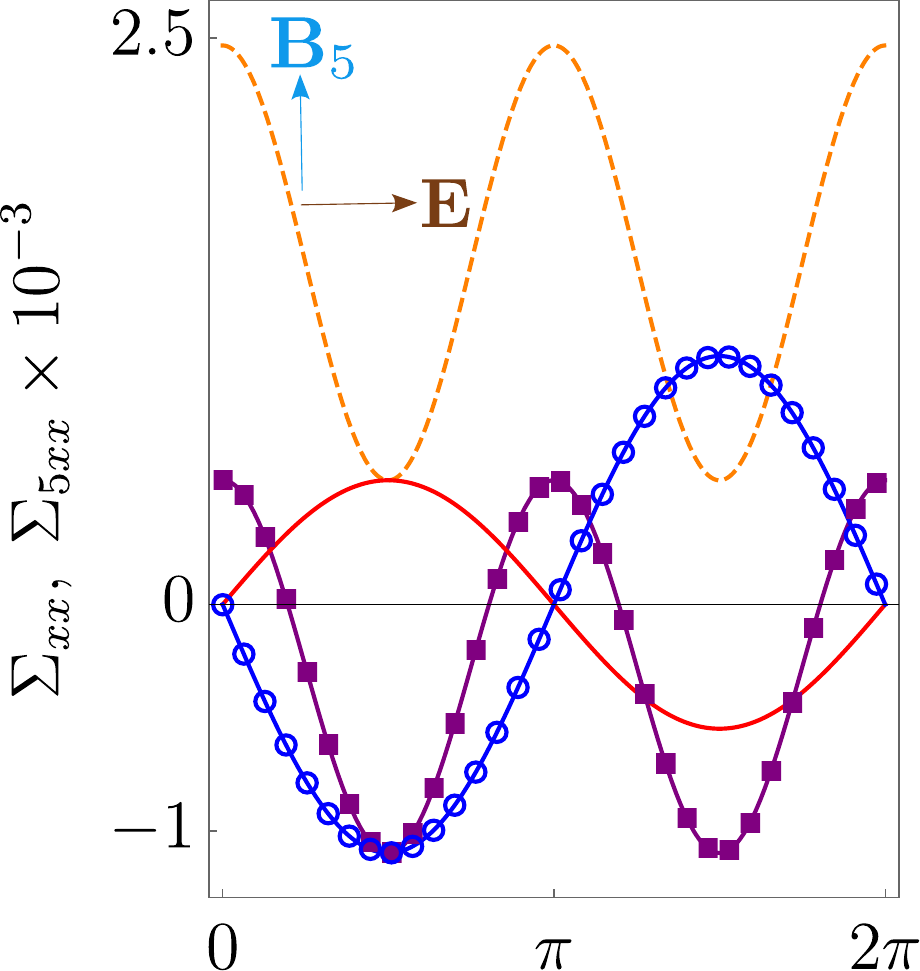} \\[6pt] \includegraphics[width=0.23\textwidth]{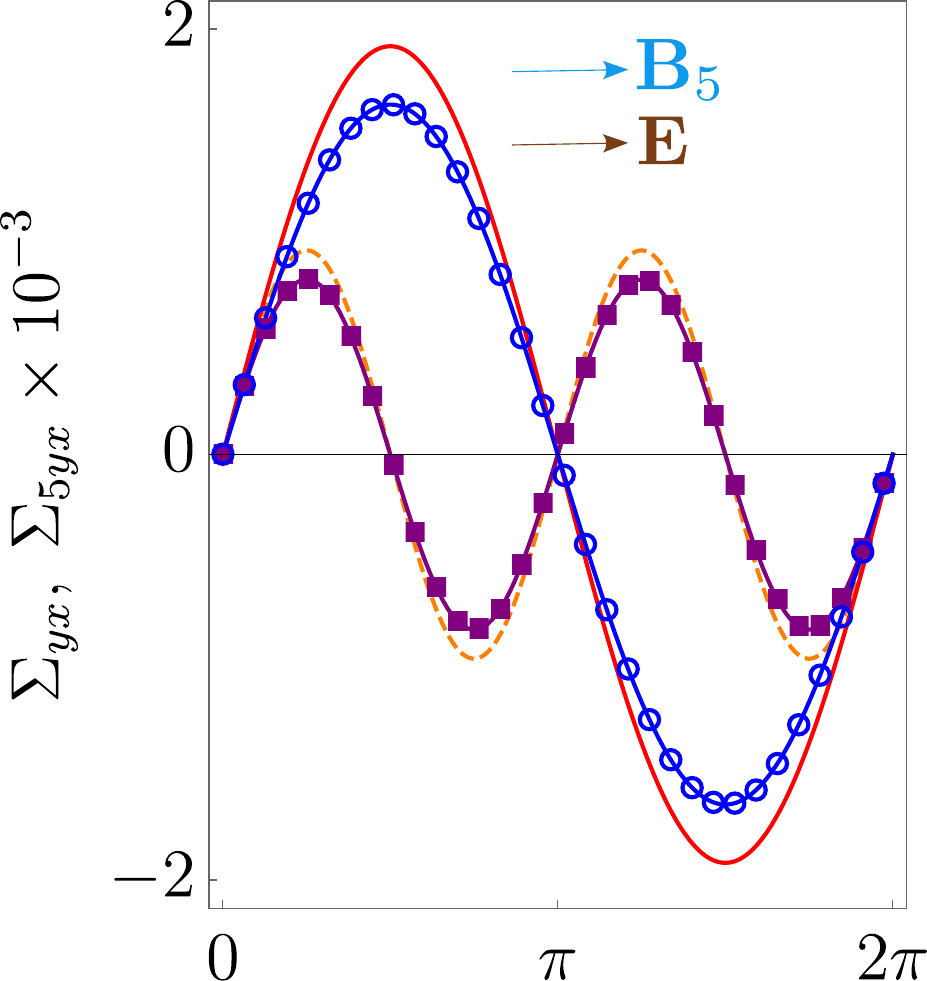}
     \includegraphics[width=0.23\textwidth]{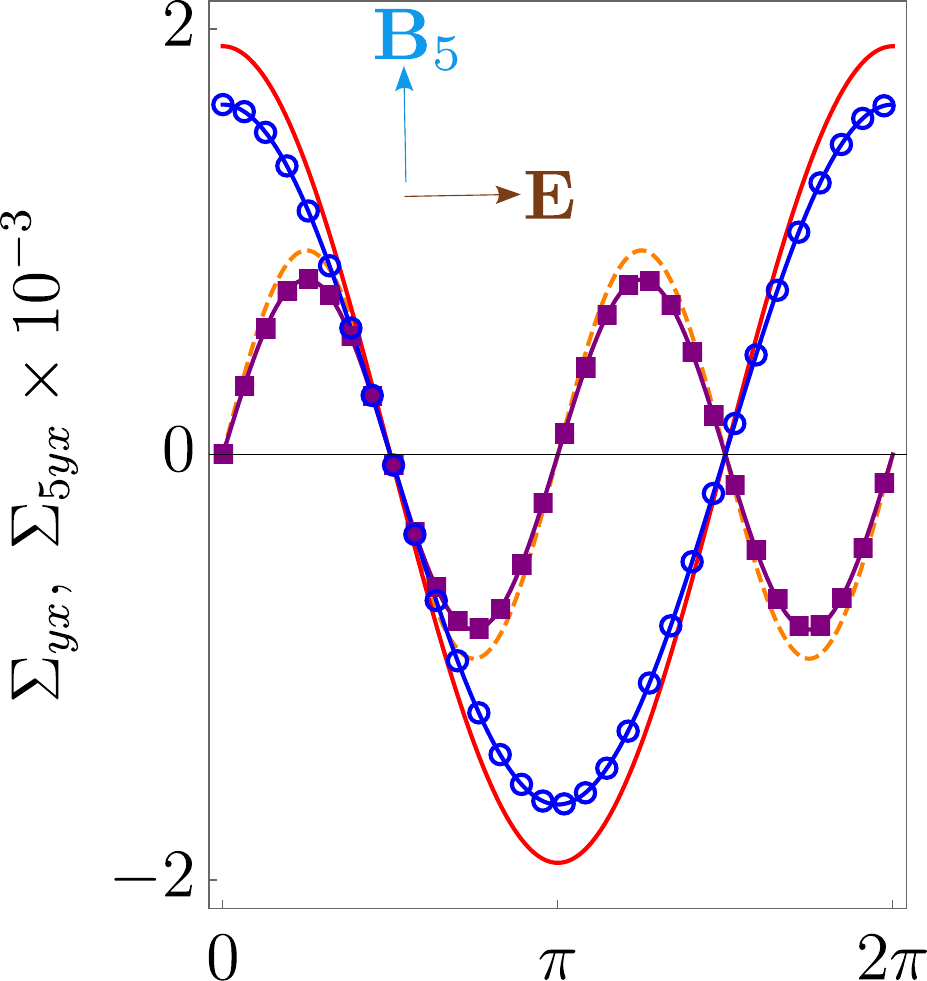}
     \caption{Longitudinal (upper panel) and planar Hall (lower panel) conductivities for $\theta _{5} = 0$ (left panel) and $\theta _{5} = \pi /2$ (right panel). The  dashed orange (continuous red) line shows the Berry curvature contribution to the total (axial) conductivities, while the purple (blue) line with square (circle) markers shows the full contribution (including the orbital magnetic moment contribution)  to the total (axial) conductivities.} \label{Conduc_T_C}
\end{figure}

In the problem at hand, using the conductivity tensor (\ref{Conductivity_Tensor}), together with the three contributions (\ref{B0_Conductivity}), (\ref{Conductivity_Tensor_Berry_Final}) and (\ref{Conductivity_Tensor_MagMoment_Final}), one can obtain general expressions for the total and chiral conductivity tensors. Assuming $b _{0 \chi}=0$ for simplicity,  and taking $T=0$K in view of the above discussion, the total conductivity becomes
\begin{align}
\sigma _{ij}  ({\boldsymbol{B}} , {\boldsymbol{B}} _{5} ) &= \sigma _{0}  \delta _{ij} + \frac{\sigma _{0}}{10 B _{0} ^{2}} \Big[  - \delta _{ij} (B^{2} + B ^{2} _{5} )  \notag \\ & \hspace{2.55cm} + 3 ( B _{i} B _{j} + B _{5i} B _{5j} )    \Big]  ,  \label{Total_conductivity}
\end{align}
where $\sigma _{0} \equiv \frac{e ^{2} \mu ^{2} \tau }{3 \pi ^{2} \hbar ^{3} v _{F} } $ is the field-independent longitudinal conductivity and $B _{0} \equiv (e \hbar ) ^{-1} (  \mu  / v _{F}  ) ^{2}$ is a characteristic magnetic field. Similarly, for the chiral conductivity we obtain
\begin{align}
\sigma  _{5ij} ({\boldsymbol{B}} , {\boldsymbol{B}} _{5} ) =  \frac{ \sigma _{0}}{10 B _{0} ^{2}} \left[ - 2 \delta _{ij} {\boldsymbol{B}} \cdot {\boldsymbol{B}} _{5}  + 3 (B _{i} B _{5j} + B _{5i} B _{j} )  \right]   .  \label{Chiral_conductivity}
\end{align}
Clearly, the chiral conductivity vanishes if either $\boldsymbol{B} $ or $ \boldsymbol{B} _{5}$ are zero. Therefore, in such case, to probe the chiral current, a pseudo-electric field is required. This is so because the axial gauge fields couple opposite chiral fermions with opposite signs. To illustrate the angular dependence of these conductivities we introduce the normalized total conductivity
\begin{align}
\Sigma _{ij} ({\boldsymbol{B}} , {\boldsymbol{B}} _{5} ) = \frac{ \sigma _{ij} ({\boldsymbol{B}} , {\boldsymbol{B}} _{5} ) - \sigma _{0}  \delta _{ij} }{ \sigma _{0}  } ,  \label{Norm_Total}
\end{align}
and similarly one can define the normalized axial conductivity as $\Sigma _{5ij} ({\boldsymbol{B}} , {\boldsymbol{B}} _{5} ) = \sigma  _{5ij} ({\boldsymbol{B}} , {\boldsymbol{B}} _{5} )  / \sigma _{0}$.  Note that expressions (\ref{Chiral_conductivity}) and (\ref{Norm_Total}) are valid for arbitrary orientations of the fields ${\boldsymbol{B}}$ and ${\boldsymbol{B}} _{5} $. Now we take the same planar configuration as before, depicted in Fig. \ref{Setup}. {To elucidate the interplay of the genuine and axial magnetic fields, }we fix the direction of the effective electric field to be along $+ \hat{\boldsymbol{e}} _{x}$, i.e. $\boldsymbol{E} _{\chi} = E _{\chi} \hat{\boldsymbol{e}} _{x}$, and rotate the magnetic field along the $xy$-plane such that it makes an angle $\theta$ with respect to the direction of the electric field, i.e. $\boldsymbol{B} = B ( \cos \theta \hat{\boldsymbol{e}} _{x} + \sin \theta \hat{\boldsymbol{e}} _{y})$. Also we take the pseudo-magnetic field in the same plane, i.e. $\boldsymbol{B} _{5} = B _{5} ( \cos \theta _{5} \hat{\boldsymbol{e}} _{x} + \sin \theta _{5} \hat{\boldsymbol{e}} _{y})$. In this case, the normalized total and axial longitudinal conductivities become
\begin{align}
\Sigma _{xx} ({\boldsymbol{B}} , {\boldsymbol{B}} _{5} ) &=  \frac{B^{2}   ( 3 \cos ^{2} \theta - 1 ) + B _{5} ^{2}  ( 3 \cos ^{2} \theta _{5} - 1 )  }{ 10 B _{0} ^{2} } , \label{ST_xx}  \\[5pt]  \Sigma  _{5xx} ({\boldsymbol{B}} , {\boldsymbol{B}} _{5} ) &=  \frac{ BB_{5} }{10 B _{0} ^{2}} \left[ 3 \cos (\theta + \theta _{5}) +  \cos (\theta - \theta _{5})    \right]  ,  \label{S5_xx}
\end{align}
respectively.  The corresponding expressions for the normalized total and axial planar Hall conductivities are
\begin{align}
\Sigma _{xy} ({\boldsymbol{B}} , {\boldsymbol{B}} _{5} ) &= \frac{3}{20 B _{0} ^{2}}   \left[ B ^{2} \sin ( 2\theta ) + B _{5} ^{2} \sin ( 2 \theta _{5} ) \right] ,    \label{ST_xy}  \\[5pt]  \Sigma  _{5xy} ({\boldsymbol{B}} , {\boldsymbol{B}} _{5} ) &=  \frac{3 B B _{5} }{10 B _{0} ^{2}} \sin ( \theta + \theta _{5} ),  \label{S5_xy}
\end{align}
respectively. We now plot these expressions. To this end, we take $B=B_{5}=0.5$ T and use typical parameters for TaAs ($v_{F} = 3 \times 10 ^{5}$ m/s and $\mu=20$ meV) which yields $B _{0} = 6.75$ T. In Fig. \ref{Conduc_T_C} we show the longitudinal (upper panel) and planar Hall (lower panel) conductivities as a function of the angle $\theta$ and fixed values $\theta _{5} = 0$ (at left) and $\theta _{5} = \pi /2$ (at right). Each panel displays both the normalized total and axial conductivities. The dashed orange (continuous red) line corresponds solely to the Berry curvature contribution to the total (axial) conductivities. The purple (blue) line with square  (circle) markers shows the full contribution (i.e. including both Berry curvature and OMM) to the total (axial) conductivities.

\begin{figure*}
\centering
\includegraphics[width=0.2\textwidth]{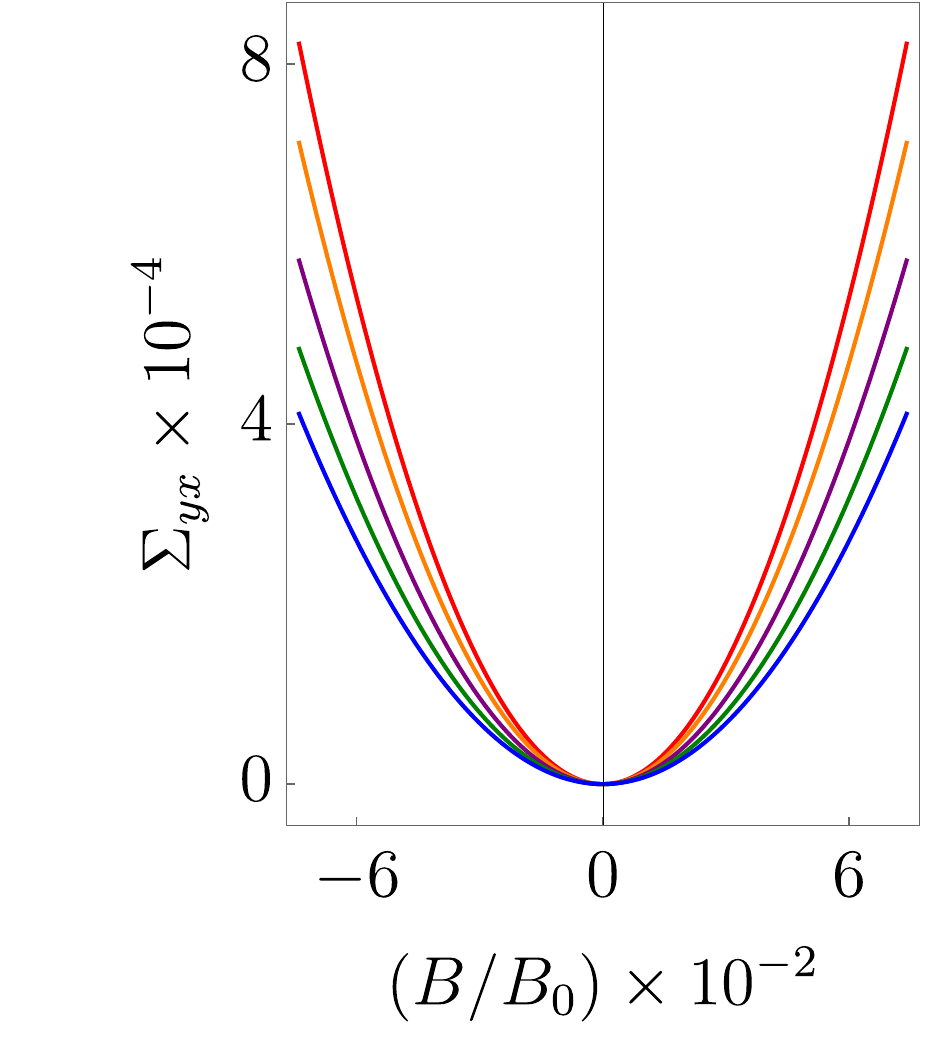} \;  \includegraphics[width=0.28\textwidth]{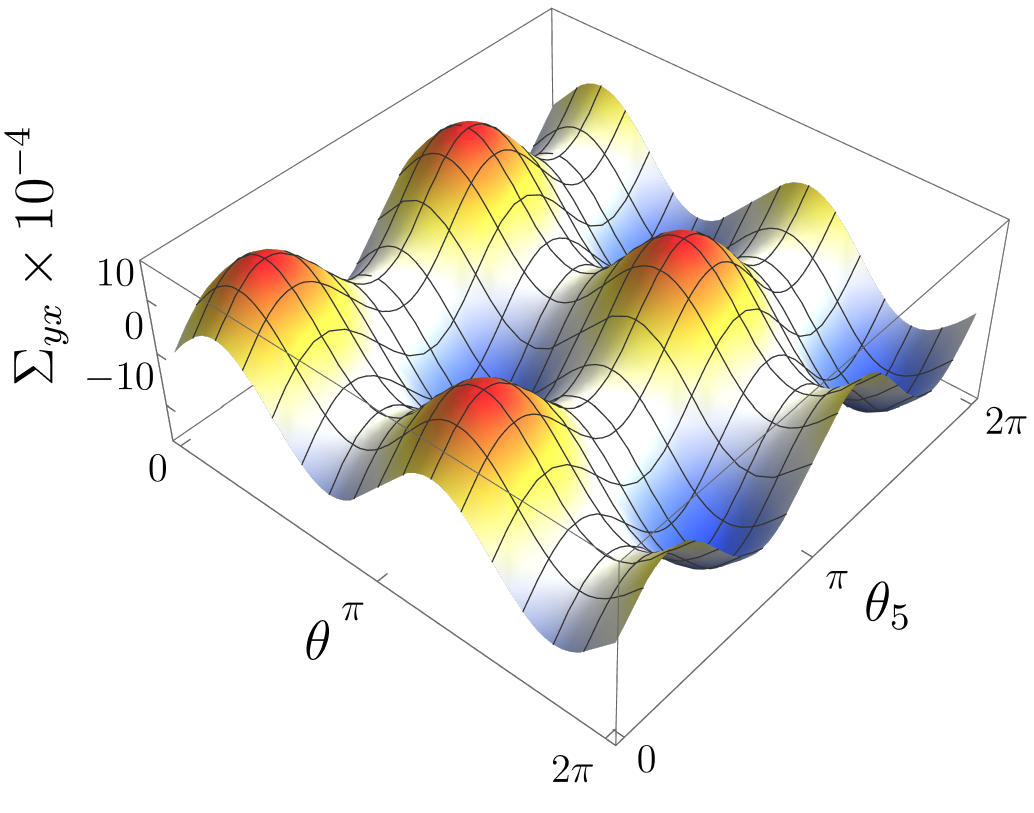} \; \includegraphics[width=0.2\textwidth]{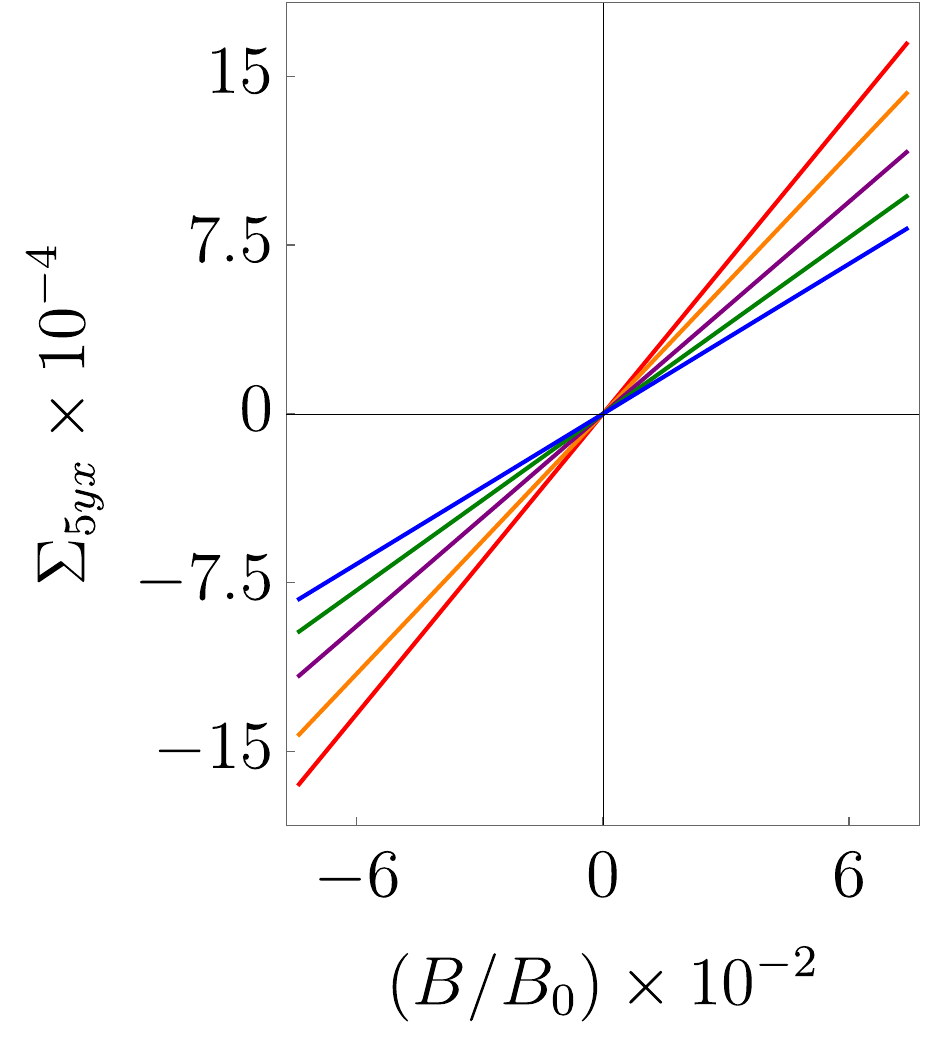}\; \includegraphics[width=0.28\textwidth]{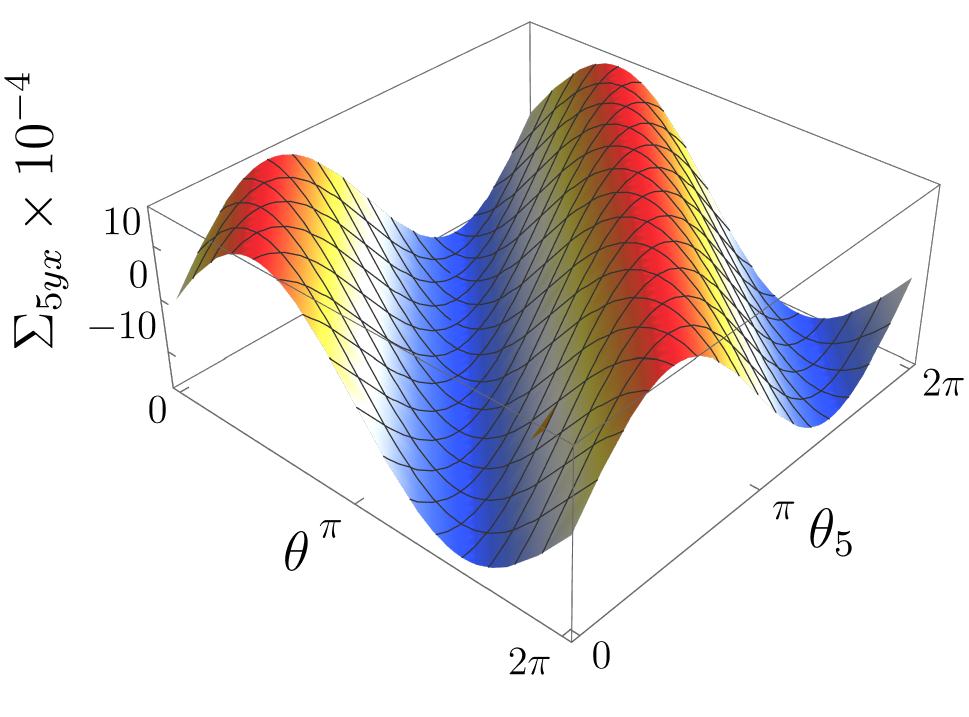} \\ (a) \hspace{4cm} (b) \hspace{4cm} (c) \hspace{4cm} (d)  
  \caption{(a) and (c) show the total and axial planar Hall conductivities, respectively, as a function of the magnetic field for $\theta _{5} = 0$, $B_{5}=0.5$ T and different values of the angle $\theta$. (b) and (d) show 3D plots of the planar and axial Hall conductivities, respectively, as a function of the angles $\theta$ and $\theta _{5}$ for $B=B_{5}=0.5$ T. } \label{QQQ}
\end{figure*}
 
The importance of the OMM is quite evident in the case of the total and axial longitudinal conductivities: the unmarked lines significantly differ from those with markers.  However, the total and axial planar Hall conductivities exhibit slight changes which can be appreciated near to some specific angles. For example, in the case of the total planar Hall conductivity $\Sigma _{yx}$, the differences are important near to $\theta ^{\ast} = n \pi / 4$, where $n=1,3,5,7$, in both situations $\theta_{5} = 0$ and $\theta _{5} = \pi /2$. For an arbitrary value of $\theta _{5}$ the position of these critical angles does not change, but the value of the normalized conductivity is shifted by $\frac{3}{20} (  B _{5} /  B _{0} ) ^{2} \sin ( 2 \theta _{5} )$. In the case of the axial conductivity, the main differences are near to $\theta ^{\ast}  = \pi /2 , \, 3 \pi / 2$ for $\theta _{5}=0$ and  near to $\theta ^{\ast}  = 0 , \,  \pi $ for $\theta _{5}= \pi / 2$. The behavior of the axial conductivity is quite the same for any value of $\theta _{5}$; the only difference is that the critical angles are shifted by $\theta _{5}$, i.e.,  they are given by $\theta ^{\ast} = \theta _{5} + n \pi / 2$.

Apart from the angular dependence of the longitudinal and planar Hall conductivities shown in Fig. \ref{Conduc_T_C}, the dependency on the magnitude of the magnetic field would also be relevant for an experimental detection of our results. In fact, as we read off from Eqs. (\ref{ST_xx})-(\ref{S5_xy}),  the total longitudinal $\Sigma _{xx}$ and transverse  $\Sigma _{xy}$ conductivities show $B ^{2}$ dependence, while the axial conductivities $\Sigma  _{5xx}$ and $\Sigma  _{5xy}$ exhibit a linear dependency on $B$. In Figs. \ref{QQQ}(a) and \ref{QQQ}(c) we plot the total and axial planar Hall conductivities, $\Sigma _{xy} ({\boldsymbol{B}} , {\boldsymbol{B}} _{5} )$ and $\Sigma _{5xy} ({\boldsymbol{B}} , {\boldsymbol{B}} _{5} )$, respectively, as a function of the (normalized) magnetic field $B/B_{0}$, for $\theta _{5} = 0$, $B_{5}=0.5$ T and different values of the angle $\theta$. In Fig. \ref{QQQ}(a) the curves correspond to angles ranging from $\theta = \pi /6$ to $\theta = \pi /2$, and they closes progressively in such order, i.e. the focal length decreases as increasing the angle $\theta$.  The vertex of the parabola is located at $\Sigma _{xy} ({\boldsymbol{0}} , {\boldsymbol{B}} _{5} ) = \frac{3}{20 B _{0} ^{2}}    B _{5} ^{2} \sin ( 2 \theta _{5} )$,  which for the chosen parameter values is zero.  Different values of the angle $\theta _{5}$ only shift the parabolas to the top for $\theta _{5} \in [0,\pi/2] \cup [\pi,3\pi/2]  $ or to the bottom for $\theta _{5} \in [\pi/2, \pi] \cup [ 3\pi/2 , 2 \pi] $.  In Fig. \ref{QQQ}(c) the straight lines crosses the origin and the slop is given by $ \frac{3 B _{5} }{10 B _{0} ^{2}} \sin ( \theta + \theta _{5} )$.  The curves correspond to angles ranging from $\theta = \pi /6$ to $\theta = \pi /2$, where the slope increases. Figures \ref{QQQ}(b) and \ref{QQQ}(d) show the total and axial planar Hall conductivities, $\Sigma _{xy}$ and $\Sigma _{5xy}$, as a function of the angles $\theta$ and $\theta _{5}$, for $B=B_{5}=0.5$ T. Red (blue) shaded regions display the maximum (minimum) of the corresponding functions.  

{Everything discussed thus far correspond, in a separate fashion, to the total and axial conductivities appearing in the general formulas (\ref{Total_Current})-(\ref{Axial_Current}).  In the presence of a genuine electric field and vanishing axial electric field, the plots presented in Figs. \ref{Conduc_T_C} and \ref{QQQ} correctly describes the behaviour of the longitudinal and planar Hall conductivities. In the opposite case, for a vanishing genuine electric field and nonzero axial electric field, the roles of the total and axial conductivities becomes interchanged, as Eqs. (\ref{Total_Current}) and (\ref{Axial_Current}) suggest. 
However, in the presence of both genuine and axial electric fields, the effects of the total $\Sigma _{ij}$ and axial $\Sigma _{5ij}$ conductivities become intertwined.  To elucidate the interplay of the electric fields, we now consider the configuration shown in Fig. \ref{Setup} in the presence of an electric field ${\boldsymbol{E}} = E _{x} \hat{{\boldsymbol{e}}} _{x}$ and an in-plane axial electric field ${\boldsymbol{E}} _{5} = E _{5} ( \cos \varphi  \hat{{\boldsymbol{e}}} _{x} + \sin \varphi  \hat{{\boldsymbol{e}}} _{y} ) $.  In this case, the total and axial currents become
\begin{align}
J _{i} / E _{x} &= \sigma _{ix}  + \epsilon _{5} ( \cos \varphi \, \sigma  _{5ix}  + \sin \varphi \, \sigma  _{5iy}  ) ,  \label{TotalE5}  \\[5pt]  J _{5i}  / E _{x}  &= \sigma _{5ix}  +  \epsilon _{5} ( \cos \varphi \, \sigma  _{ix}  + \sin \varphi \, \sigma  _{iy}  )  ,  \label{Total5E5}
\end{align}
\begin{figure}
    \centering
	\includegraphics[width=0.22\textwidth]{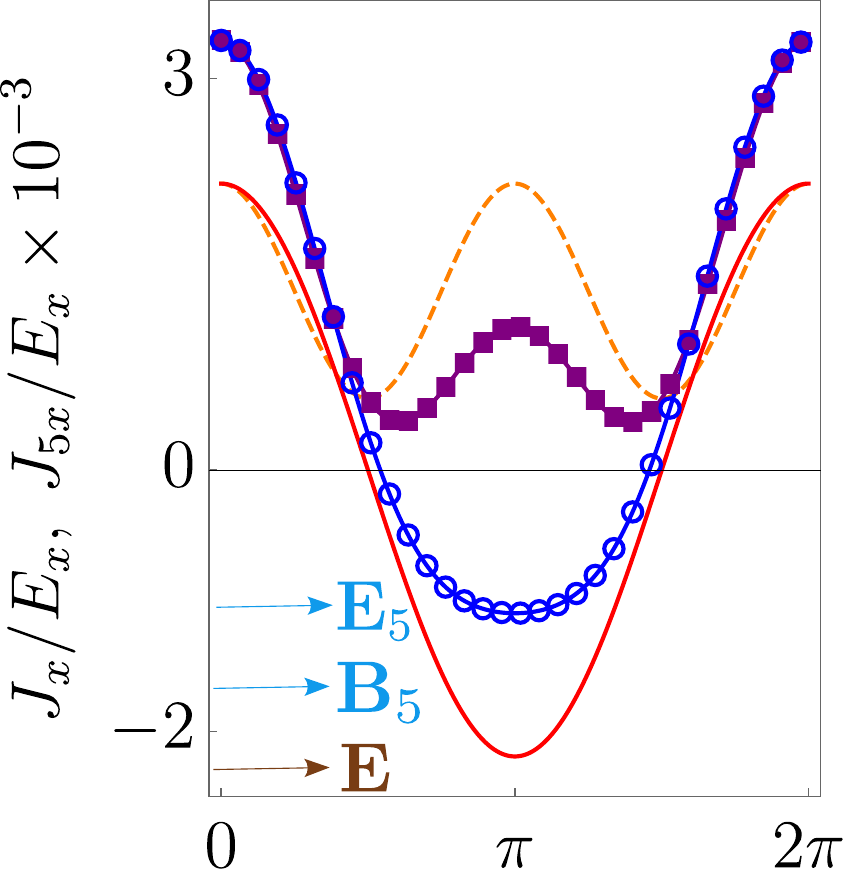} 
	\includegraphics[width=0.22\textwidth]{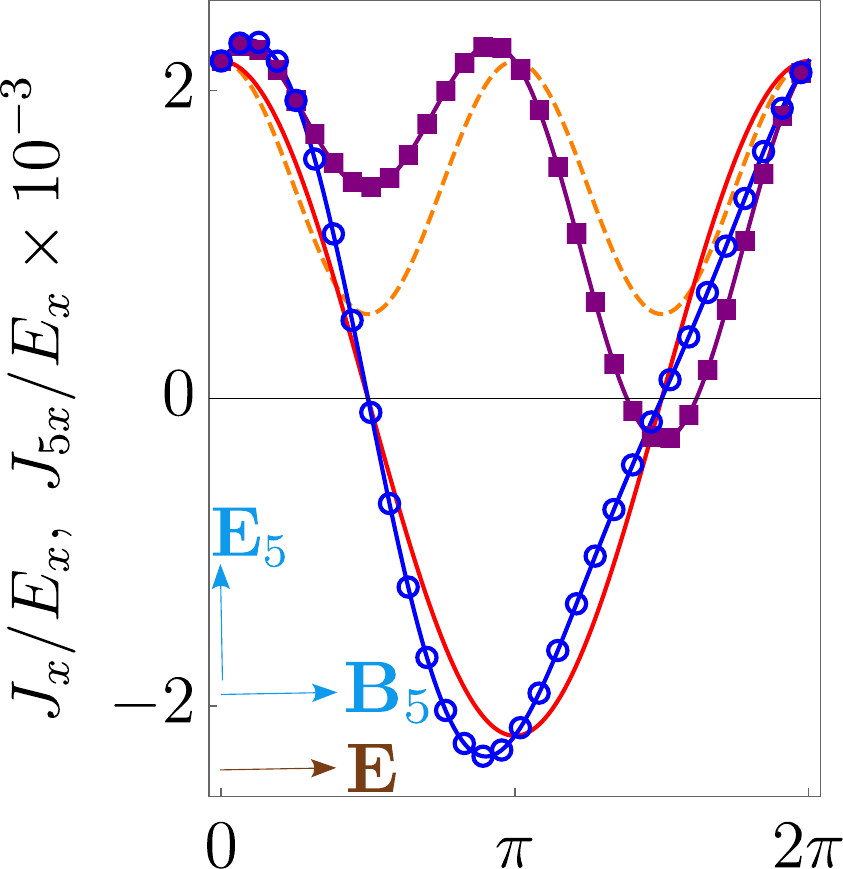}
	\\[6pt] \includegraphics[width=0.22\textwidth]{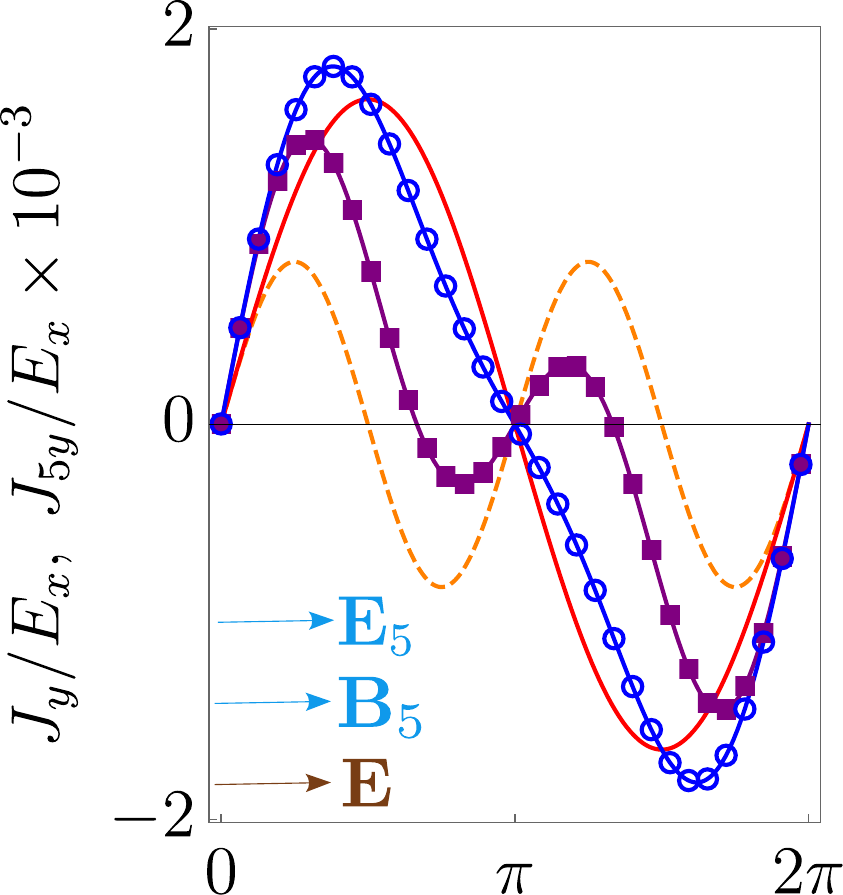}
	\includegraphics[width=0.23\textwidth]{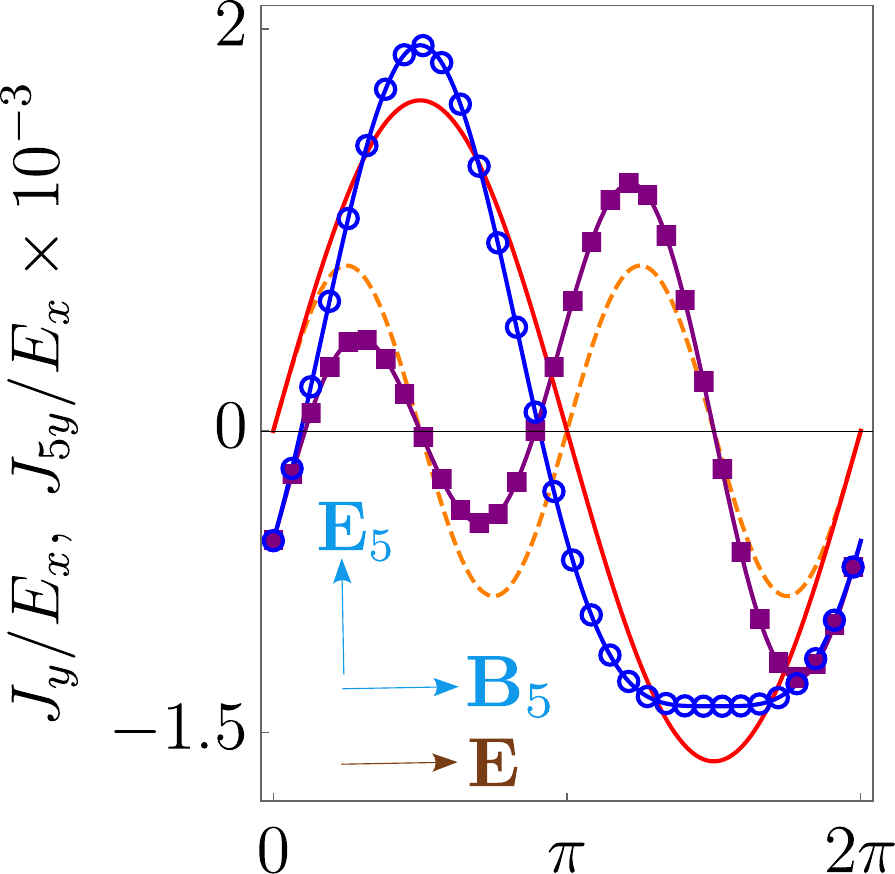}
     \caption{Angular dependence of the longitudinal (upper panel) and planar Hall (lower panel) currents for ${\boldsymbol{E}} = E _{x} \hat{{\boldsymbol{e}}} _{x}$, $\theta _{5} = 0$, $\varphi = 0$ (left panel) and $\varphi = \pi /2$ (right panel). The  dashed orange (continuous red) line shows the currents for $E_{5}=0$, while the purple (blue) line with square (circle) markers shows the full currents including the axial electric field.   } \label{Conduc_Mixed}
\end{figure}

where $\epsilon _{5} \equiv E _{5} / E _{x}$ is the ratio between the axial and genuine electric field, and $\sigma _{ij}$ and $\sigma _{5ij}$ are the conductivity tensors given by Eqs. (\ref{Total_conductivity}) and (\ref{Chiral_conductivity}), respectively. To plot these functions we take appropriate values for the genuine and axial fields. Genuine electromagnetic fields are extensively controlled in experiments. However, strain-induced fields are more subtle. For example, for a WSM in the presence of torsion, the maximum attainable axial magnetic field is $B _{5} \approx 0.5$T \cite{PhysRevX.6.041021}. For simplicity we also take $B=0.5$T. An estimate for the axial electric field can be obtained from the upper limit $E _{5} \leq v _{F} B _{5}$, which is required to circumvent the collapsing of Landau levels in WSMs \cite{PhysRevB.96.081110}. In TaAs we find $E _{5} \leq 1.5 \times 10 ^{5}$V/m, and then here we take $E _{5} = 2 \times 10 ^{4}$V/m,  which vastly fulfils the condition. In the next Section we deepen on the origin of strain-induced pseudofields. If we take a moderate electric field of strength  $E _{x} = 2 \times 10 ^{4}$V/m we obtain $\epsilon _{5} = 0.5$.  In the following, to better understand the interplay of the axial electric field, we fix the axial magnetic field pointing parallel to the electric field, i.e.  ${\boldsymbol{B}} = B \hat{{\boldsymbol{e}}} _{x}$, and explore the angular dependence of the full currents (Berry curvature + OMM contributions) as a function of the angles $\theta$ (defined by the magnetic field) and $\varphi$. In the upper panel of Fig. \ref{Conduc_Mixed} we plot the longitudinal total $J _{x}$ and axial $J _{5x}$ currents (in units of $E_{x}$) as a function of the angle $\theta$ for $\theta _{5} = 0$ and two different values of $\varphi$, namely, $\varphi = 0$ (left panel) and $\varphi = \pi/2$ (right panel). The lower panel displays the total $J _{y}$ and axial $J _{5y}$ planar Hall currents (in units of $E_{x}$). {The dashed orange and continuous red lines correspond to the currents in the absence of the axial electric field, while the purple and blue lines with square and circle markers correspond to the currents in the presence of $E _{5}$. We observe that for an axial electric field in parallel configuration (i.e. with $\varphi = 0$), the longitudinal total and axial currents behave in a similar manner than in the absence of $E_{5}$, changing sign approximately around the same angles. However, in a perpendicular configuration (i.e. with $\varphi = \pi /2$), the angular dependence of the current is slightly different, with the position of the extrema shifted (indeed, the shifts can be directly calculated from Eqs. (\ref{TotalE5})-(\ref{Total5E5})). We highlight the fact that the total conductivity (purple line with square markers) reverses its sign near to $\theta = 3\pi / 2$, and this is a direct consequence of the nonzero axial electric field.  Therefore, this is a direct signature of the effects of the $E_{5}$ upon the longitudinal conductivities. The case of the planar Hall conductivities is similar. In parallel configuration ($\varphi = 0$), the planar Hall current both in the presence and in the absence of the axial electric field display the same angular dependence, flipping sign around the same angles, with the extrema slightly shifted and with the amplitude controlled by the field strength $E_{5}$.  In perpendicular configuration, there are two aspects that could be highlighted. On the one hand, in the case $E_{5} = 0$ there are three angles at which the current vanishes, i.e. $\theta = 0, \pi, 2 \pi$, since $J_{y} \sim 3 \sigma _{0} E _{x} \sin \theta \cos \theta$; however, in the presence of an axial electric field the planar Hall current is of the form $J _{y}  \sim \sigma _{0} E _{x}  ( 3 \sin \theta - 2 \epsilon _{5} ) \cos \theta  $, thus implying that there is only one critical angle at which the current vanishes, given by $\theta ^{\ast} = \arcsin (2 \epsilon _{5} / 3)$, and take the constant value $J _{y} \sim -  2 \sigma _{0} \epsilon _{5} E _{x} $ at $\theta = 0 , 2 \pi$. On the other hand, as the lower-right panel of Fig. \ref{Conduc_Mixed} shows, in the presence of the axial electric field the periodicity of the current in the angle $\theta$ is broken, which becomes clear from the previous equations for the currents with and without $E _{5}$. These are distinguishing features of the presence of the axial electric field. All these conclusions are also supported by the 3D plots presented in Fig. \ref{Conduc_3D}, which display the total and axial planar Hall currents (in units of $E_{x}$) as a function of the angles $\theta$ and $\varphi$. Red (blue) shaded regions display the maximum (minimum) of the corresponding functions.  

  }

\begin{figure}
    \centering
	\includegraphics[width=0.33\textwidth]{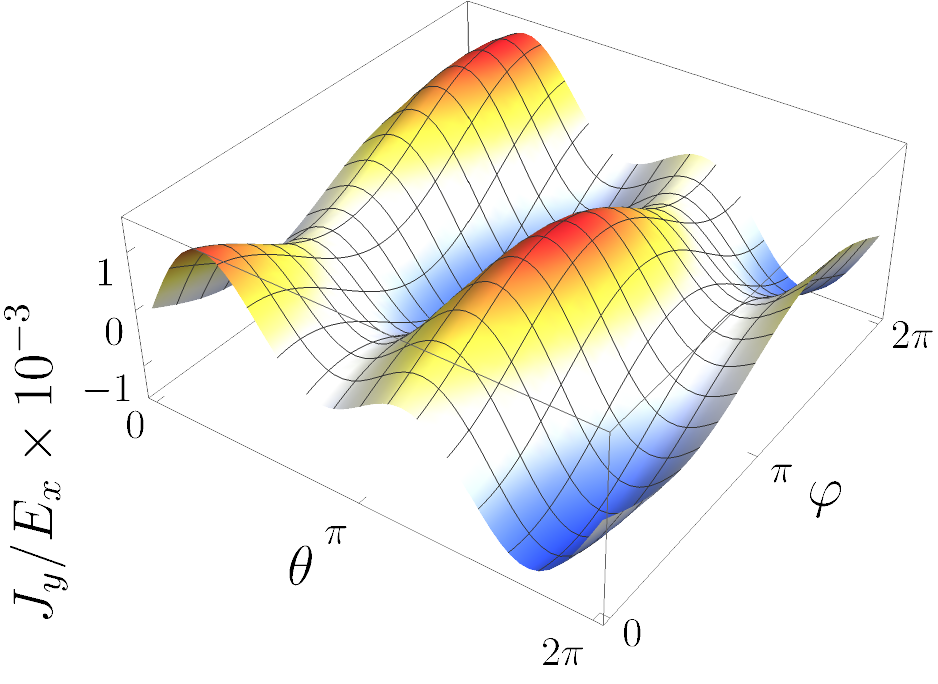} 
	\includegraphics[width=0.33\textwidth]{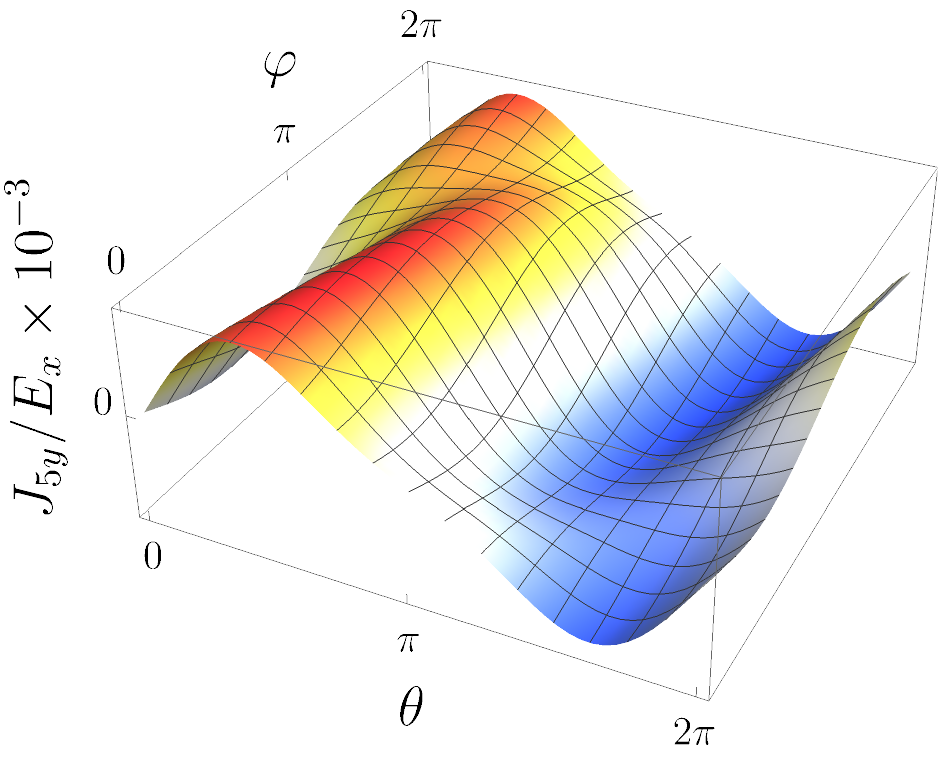}
     \caption{3D plots of the total and axial planar Hall current (in units of $E_{x}$) as a function of the angles $\theta$ (defined by the magnetic field) and $\varphi$ (defined by the axial electric field) } \label{Conduc_3D}
\end{figure}

}

As pointed out, anomaly related transport phenomena have attracted great attention in condensed matter physics. For example, in graphene, fermionic excitations near the Dirac cones are described by a (2+1)-dimensional quantum field theory exhibiting the parity anomaly.  This gives rise to the valley Hall effect, observed experimentally \cite{doi:10.1126/science.1254966, 10.1038/nphys3551} and extensively explored due to possible applications in valleytronics  \cite{10.1038/nphys547}. On the same footing, Weyl semimetals provide an electronic route for realizing the chiral anomaly in condensed matter. The chiral anomaly induces a number of novel phenomena in WSMs, including the chiral magnetic effect \cite{PhysRevD.78.074033} that manifests itself through the negative longitudinal magnetoresistance \cite{PhysRevB.88.104412, PhysRevLett.113.247203}, and it was confirmed in transport experiments in the TaAs family \cite{PhysRevX.5.031023, 10.1038/ncomms10735}.  The planar Hall effect has  been ascribed also to the chiral anomaly; however, as shown in this paper, there are other contributions that affect the PHE in a similar order of magnitude than the Berry curvature contribution, namely, the orbital magnetic moment of charge carriers.

WSMs in the presence of pseudofields give unique opportunities to  probe the different contributions to the covariant anomaly equations (\ref{Chiral_Anomaly})-(\ref{Charge_Anomaly}).  Along this line,  our results provide a testing ground for the chiral and charge anomalies by means of the planar Hall effect. Experimentally,  pseudofields are induced by applying strain on the crystal.  For example, a pseudomagnetic field  ${\boldsymbol{B}} _{5}$ can be created by applying a static torsion or bending the sample. A nonzero pseudoelectric field ${\boldsymbol{E}} _{5}$ is generated, for instance, by dynamically stretching or compressing the sample.  Therefore,  properly combining genuine and pseudo-electromagnetic fields, we are able to test the four terms in the right-hand side of Eqs. (\ref{Chiral_Anomaly}) and (\ref{Charge_Anomaly}). On the one hand, the two terms in the chiral anomaly equation (\ref{Chiral_Anomaly}),  ${\boldsymbol{E}} \cdot {\boldsymbol{B}}$ and ${\boldsymbol{E}} _{5} \cdot {\boldsymbol{B}} _{5}$, can be tested by using nonorthogonal electromagnetic fields or pseudo-electromagnetic fields, respectively. The latter can be achieved by a simultaneous application of torsion and time-dependent unidirectional strain.  Physically, in both cases, this can be understood  as pumping of charge from one node to the other.  On the other hand, the two terms in the charge anomaly equation (\ref{Charge_Anomaly}),  ${\boldsymbol{E}} \cdot {\boldsymbol{B}} _{5}$ and ${\boldsymbol{E}} _{5} \cdot {\boldsymbol{B}} $,  can be tested by the combined application of electromagnetic and pseudo-electromagnetic fields.  These nonconserving charge terms should be interpreted with caution, since in a real solid, charge is strictly conserved. Indeed, they can be understood as pumping of charge between the bulk and the boundary of the system \cite{PhysRevB.87.235306, PhysRevB.89.081407, PhysRevX.6.041021}. This implies that additional currents must exist in the system to restore charge conservation. This problem is solved by adding Bardeen polynomials \cite{PhysRevB.99.140201, BARDEEN1984421, PhysRevLett.118.127601}, which in the present case, corresponds to the topological Chern-Simons charge and current densities, in the order given by $\rho _{\mbox{\scriptsize CS}} =  - \frac{e ^{2}}{2 \pi ^{2} \hbar ^{2} }   {\boldsymbol{b}} \cdot {\boldsymbol{B}}$ and ${\boldsymbol{J}} _{\mbox{\scriptsize CS}} =   \frac{e ^{2}}{2 \pi ^{2} \hbar ^{2} } ( -  b_{0} {\boldsymbol{B}} +   {\boldsymbol{b}} \times {\boldsymbol{E}} ) $, being $2{\boldsymbol{b}}$ and $2b_{0}$ the separation between the Weyl nodes in momentum and energy, respectively, as depicted in Fig. \ref{Fig_Cones}. To understand this result physically, we must recall that pseudo-fields are induced by inhomogeneous strain, which results in a  position and time dependent separation between the nodes in momentum and energy, i.e. ${\boldsymbol{b}} ({\boldsymbol{r}},t)$ and $b _{0} ({\boldsymbol{r}},t)$, such that the emerging pseudo-fields are determined by ${\boldsymbol{B}} _{5} =  \nabla \times {\boldsymbol{b}}$ and ${\boldsymbol{E}} _{5} =  - \nabla b _{0} - \partial _{t} {\boldsymbol{b}}$.  It is now straightforward to check that the Chern-Simons 4-current $J ^{\mu} _{\mbox{\scriptsize CS}} = ( c \rho _{\mbox{\scriptsize CS}} ,  {\boldsymbol{J}} _{\mbox{\scriptsize CS}} ) $ satisfies $\partial _{\mu} J ^{\mu} _{\mbox{\scriptsize CS}} = \frac{e ^{2}}{2 \pi ^{2} \hbar ^{2} } ( \boldsymbol{E} \cdot \boldsymbol{B} _{5} + \boldsymbol{E} _{5} \cdot \boldsymbol{B} ) $. This fact suggests the definition of consistent total and axial currents $J _{\mbox{\scriptsize cons}} ^{\mu} = J  ^{\mu} +  J ^{\mu} _{\mbox{\scriptsize CS}}$ and $J _{5\mbox{\scriptsize cons}} ^{\mu} = J  ^{\mu} _{5} +  J ^{\mu} _{5 \mbox{\scriptsize CS}}$,  respectively, restoring charge conservation.  In this manner, the consistent version of the anomaly equations (\ref{Chiral_Anomaly}) and (\ref{Charge_Anomaly}) are
\begin{align}
\partial _{\mu} J ^{\mu} _{5 \mbox{\scriptsize cons}} &= \frac{e^{2}}{2 \pi ^{2} \hbar ^{2}} ( \boldsymbol{E} \cdot \boldsymbol{B} + \frac{1}{3} \boldsymbol{E} _{5} \cdot \boldsymbol{B} _{5} ) ,   \label{Chiral_Consistent_Anomaly} \\  \partial _{\mu} J ^{\mu} _{\mbox{\scriptsize cons}}  &= 0 ,  \label{Charge_Consistent_Anomaly}
\end{align}
respectively. Therefore, charge nonconservation stated in the anomaly equations (\ref{Chiral_Anomaly}) and (\ref{Charge_Anomaly}) not to worry us.  In the next Section we will back to this point in particular examples.

\section{Strain-induced nonlinear transport phenomena in WSMs} \label{strain_section}

So far we have discussed in general the effects of pseudo-fields upon nonlinear transport phenomena in Weyl semimetals and remarked in passing that they emerge due to position and time dependent deformations.  In this Section we deepen on the physical origin of the pseudo-fields and apply our results to particular strain configurations.

\begin{figure}
    \centering
    \includegraphics[width=0.4\textwidth]{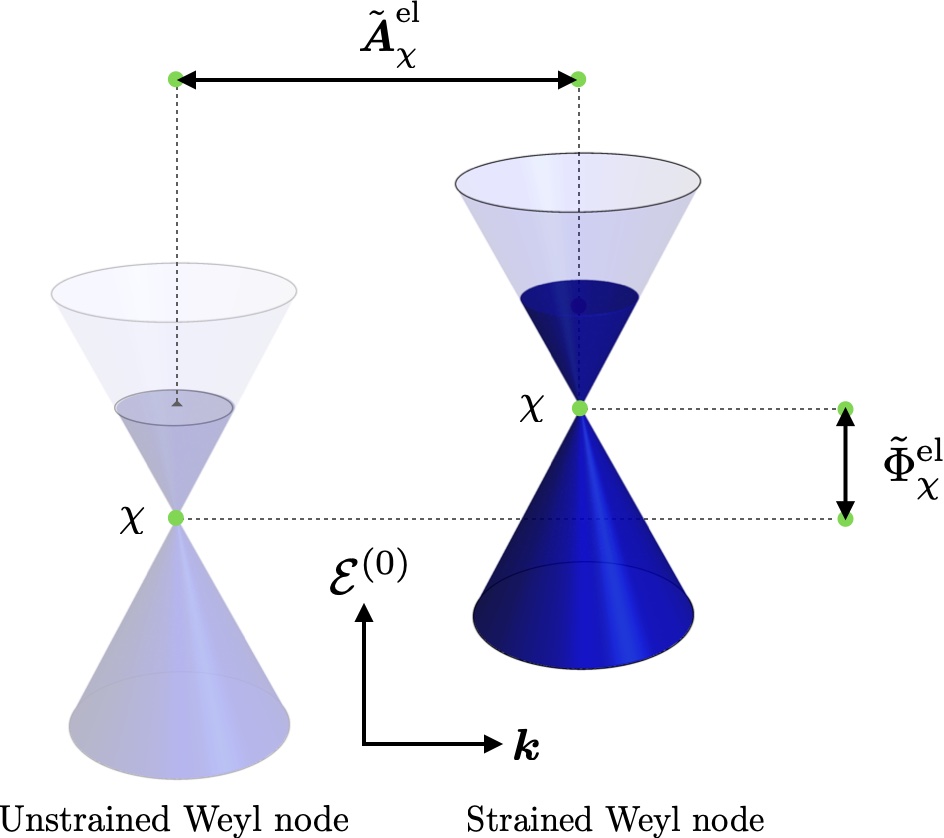}
    \caption{Strain-induced shift of a single Weyl node in momentum and energy, defining the pseudo-gauge potentials $\tilde{{\boldsymbol{A}}} _{\chi} ^{\mbox{\scriptsize el} }$ and $\tilde{\Phi} _{\chi} ^{\mbox{\scriptsize el} }$, respectively. } \label{Fig_Cone_strained}
\end{figure}

In a strained material, the modifications in the hoping parameters (between atomic orbitals) and on-site energies are determined by the components of the strain-tensor $u_{ij} = \tfrac{1}{2} \left( \partial u _{i} / \partial x ^{j} + \partial u _{j} / \partial x ^{i}  \right)$,  where $u _{i}$ is the local displacement vector of the strained lattice. In Weyl semimetals the deformation of the crystal lattice shifts the Weyl nodes in momentum and energy, as shown in Fig. \ref{Fig_Cone_strained}, and such node shifts can be described in terms of pseudo-gauge potentials $\tilde{{\boldsymbol{A}}} _{\chi} ^{\mbox{\scriptsize el} }  ({\boldsymbol{r}},t)$ and $\tilde{\Phi} _{\chi} ^{\mbox{\scriptsize el} }  ({\boldsymbol{r}},t)$ \cite{PhysRevLett.115.177202, PhysRevLett.115.177202}.  These couple to the electronic degrees of freedom as the electromagnetic vector and scalar potentials do.  Therefore, the effects of strain on a single Weyl node of chirality $\chi$, as depicted in Fig. \ref{Fig_Cone_strained}, are captured by the Hamiltonian
\begin{align}
\hat{H} _{\chi} ({\boldsymbol{k}}) = \chi  v _{F} \boldsymbol{\sigma} \cdot \big( \hbar {\boldsymbol{k}} + \tilde{{\boldsymbol{A}}} _{\chi} ^{\mbox{\scriptsize el} }   \big) + \tilde{\Phi} _{\chi} ^{\mbox{\scriptsize el} }  + b _{0 \chi} ,  \label{Hamiltonian-Strained}
\end{align}
where the pseudo-gauge potentials are expressed in terms of the strain-tensor as $\tilde{A} _{\chi i} ^{\mbox{\scriptsize el} } = h _{ijk} u _{jk}$ and $\tilde{\Phi} _{\chi} ^{\mbox{\scriptsize el} } = g _{ij} u _{ij}$.  Here, $h _{ijk}$ is the Weyl node shift per unit strain and $g _{ij}$ is the energy shift per unit strain, which have to be computed with a microscopic model (e.g. tight-binding, \textit{ab initio}) or determined by experiments \cite{PhysRevB.106.075125}.  These tensors contain all material details, such as anisotropy and elastic parameters. The corresponding pseudo-fields $\tilde{\boldsymbol{E}} _{\chi} ^{\mbox{\scriptsize el} } = - \nabla \tilde{\Phi} _{\chi} ^{\mbox{\scriptsize el} } - \partial _{t} \tilde{{\boldsymbol{A}}} _{\chi} ^{\mbox{\scriptsize el} } $ and $\tilde{\boldsymbol{B}} _{\chi} ^{\mbox{\scriptsize el} } = \nabla \times \tilde{{\boldsymbol{A}}} _{\chi}^{\mbox{\scriptsize el} } $ may couple opposite chiral fermions with opposite signs, in much the same fashion than the case of strained graphene, in which pseudo-fields couple to the Dirac fermions oppositely in the two valleys.  To account for the two possibilities, that pseudo-fields couple axially or not to the Weyl nodes, we introduce the notation $\tilde{\boldsymbol{E}} _{\chi} ^{\mbox{\scriptsize el} }  = \boldsymbol{\mathcal{E}} +  \chi \boldsymbol{E} _{5} $ and $\tilde{\boldsymbol{B}} _{\chi} ^{\mbox{\scriptsize el} }  = \boldsymbol{\mathcal{B}} + \chi \boldsymbol{B} _{5} $, where $\boldsymbol{\mathcal{E}}$ and $\boldsymbol{\mathcal{B}}$ are strain-induced pseudo-fields which couple to the Weyl nodes in the same manner as electromagnetic fields do, while $ \boldsymbol{E} _{5}$ and $\boldsymbol{B} _{5}$ are the axial parts of the pseudo-fields, which couple opposite chiral fermions with opposite signs.  All these fields are fully determined by the strain-tensor.  Note that our results of the previous Section holds in this case; however, the elastic (non-axial) electric and magnetic fields should be considered as genuine electromagnetic fields.

In order to probe strain-induced nonlinear phenomena in Weyl semimetals, we have to consider suitable non-uniform strain tensors.  In the following we shall discuss two interesting cases with experimental possibilities. To be precise, we study the effects of position-dependent strain tensors: (i) bending the WSM into a circular arc and (ii) twisting a wire-shaped WSM. These configurations produce uniform pseudo-fields, such that the pseudo-magnetic field couples axially to the Weyl fermions, and the pseudo-electric field is non-axial and then couples to the Weyl fermions in the usual manner \cite{PhysRevB.106.075125}.  With the help of an additional magnetic field, these can be used to test the covariant anomaly equations.

\subsection{Bending of WSM thin films}

Films and wires realizations of Weyl semimetals are excellent probes to test strain-induced phenomena. In fact, the best experimentally accessible geometry of applied strain is obtained by bending thin films of WSMs, which is a 3D generalization of the configuration suggested for graphene sheets in Ref.  \cite{PhysRevB.81.035408}.  As we shall see, bending of WSM thin films is an excellent test bed for the covariant anomaly equations.

\begin{figure}
    \centering
    \includegraphics[width=0.4\textwidth]{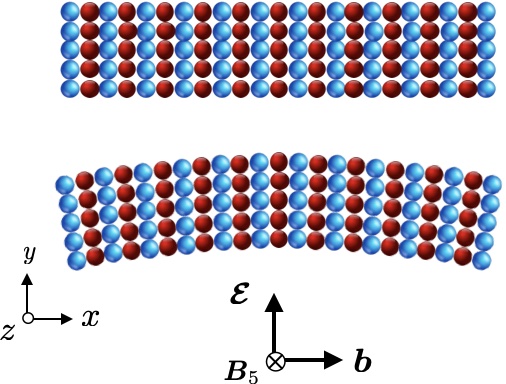}
    \caption{Sketch of the bending geometry that generates pseudo-fields. The applied strain bends the original configuration (upper panel) into a circular arc in the $x$-$y$ plane (lower panel).} \label{Fig_strained_WSM}
\end{figure}

\begin{figure}
    \centering
    \includegraphics[width=0.52\textwidth]{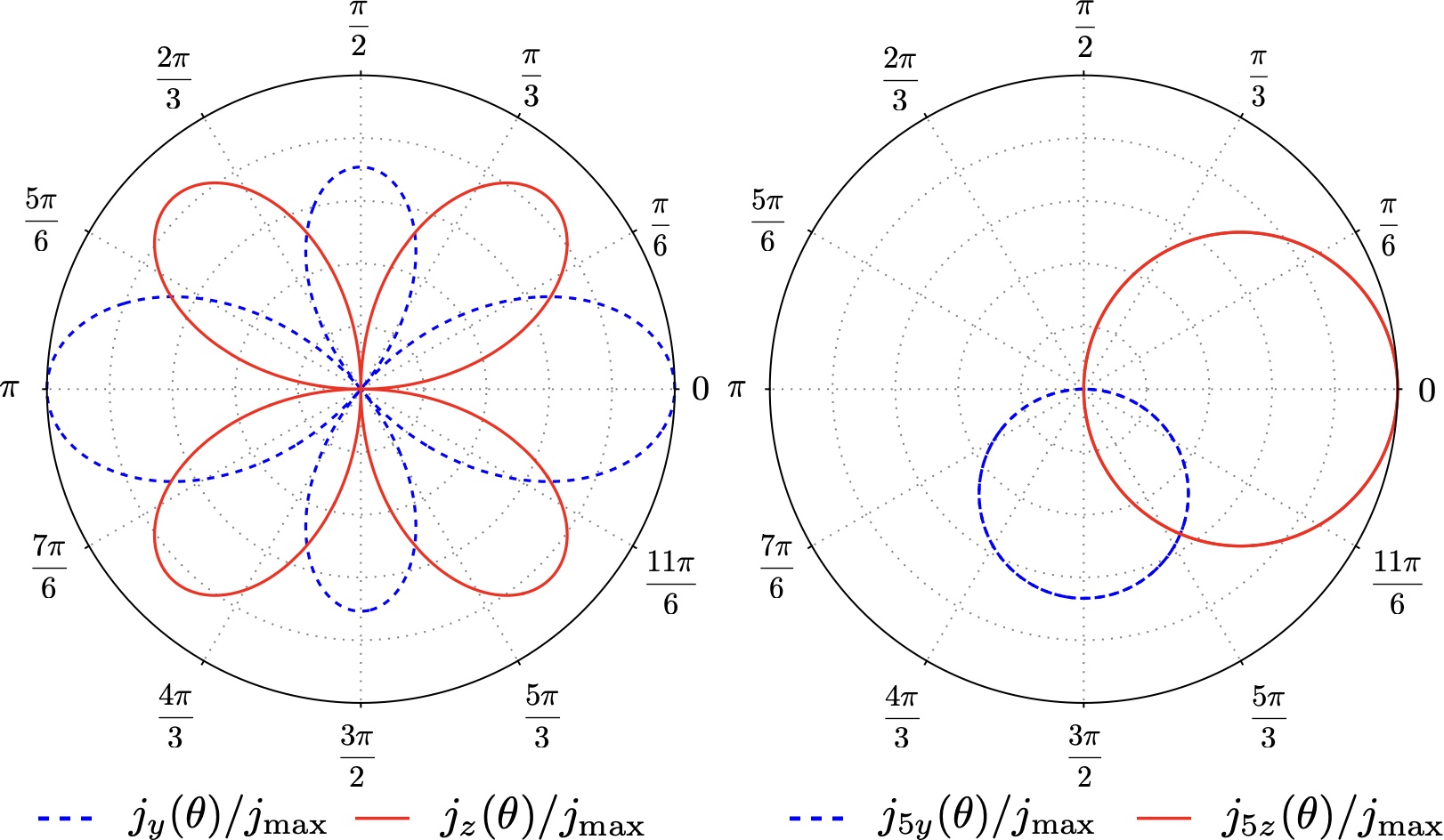}
     \caption{Angle dependence of the (normalized) total ${\boldsymbol{j}}$ (at left) and axial ${\boldsymbol{j}} _{5}$ (at right) currents for a WSM strained into a circular arc in the $x$-$y$ plane, as depicted in Fig. \ref{Fig_strained_WSM}. The blue dashed lines show the $y$-components, and the red continuous line are the $z$-components. } \label{Ang_Depend}
\end{figure}

Let us consider a rectangular lattice model of a WSM, as depicted in the upper panel of Fig.  \ref{Fig_strained_WSM}, with two nodes separated by a distance $b$ in the $k_{x}$ direction. Bending the system into a circular arc in the $x$-$y$ plane, as sketched in the lower panel of Fig.  \ref{Fig_strained_WSM},  is described by the deformation \cite{PhysRevB.95.041201}:
\begin{align}
u _{x} &= u _{0} (2xy + C x) , \\ u _{y} &= u _{0} \left[ - x ^{2} - D y(y + C ) \right]  , \\ u _{z} &= 0 ,
\end{align}
where $u_{0}$, $C$ and $D$ are constants that depend on the material. This yields the axial vector potential $A_{5x} = u _{xx} b = u _{0} (2y+C)b$, i.e. it couples to the Weyl nodes with opposite signs. The corresponding axial pseudo-magnetic field is then $B_{5z} = -2u _{0}b$. On the other hand, this strain configuration produces also a scalar (deformation) potential $\tilde{\Phi} _{\chi} = u _{xx} + u _{yy} = u _{0}(1-D)(2y+C)$, which produces the elastic (i.e. non-axial) pseudo-electric field $\mathcal{E} _{y}=-2u_{0}(1-D)g$, perpendicular to the pseudo-magnetic field, as shown in Fig.  \ref{Fig_strained_WSM}. The orthogonality between the strain-induced pseudo-fields makes this configuration appropriate to investigate the planar Hall effect in the presence of an external magnetic field in the $y$-$z$ plane, i.e. ${\boldsymbol{B}} = B (\cos \theta \hat{{\boldsymbol{e}}} _{y} + \sin \theta \hat{{\boldsymbol{e}}} _{z} ) $, coplanar to the pseudo-fields. Using equations (\ref{Total_Current}) and (\ref{Axial_Current}) one determines directly the total and axial currents.  Normalizing the currents by $\sigma _{0} \mathcal{E} _{y}$ one finds the following expressions for the total ${\boldsymbol{j}}$ and axial ${\boldsymbol{j}} _{5}$ currents
\begin{align}
{\boldsymbol{j}} (\theta ) &= \frac{1}{10}  \left( \begin{array}{c} 0 \\    \tilde{B} ^{2} (3  \cos ^{2} \theta  - 1) - \tilde{B} ^{2} _{5}   \\ 3 \tilde{B} ^{2} \sin \theta \cos \theta     \end{array} \right) , \\ {\boldsymbol{j}} _{5} (\theta )  &= \frac{  \tilde{B} _{5} \tilde{B}  }{10}  \left( \begin{array}{c} 0 \\ -  2 \sin \theta \\ 3 \cos \theta  \end{array} \right)
\end{align}
where we have subtracted the field-independent current from the total current ${\boldsymbol{j}}$. Here, $\tilde{B} = B/B_{0}$ and $\tilde{B} _{5}= B_{5}/B_{0}$.  Interestingly, these are planar Hall currents., i.e. they are in the $y$-$z$ plane, where the pseudo-fields and the magnetic field lie. The angular dependences of the components of ${\boldsymbol{j}}$ and ${\boldsymbol{j}} _{5}$ are plotted in Fig.  \ref{Ang_Depend}. We have taken $B = 1$T, $B _{5}=0.5$T, $B _{0}=6.75$T (characteristic of TaAs) and normalized the currents by its maximum values $j _{\mbox{\scriptsize max}} = j _{y}(\pi)$ and $j _{5 \mbox{\scriptsize max}} = j _{5z}(\pi/2)$. The left panel shows the total current components: the blue dashed  line corresponds to $j _{y} (\theta) / j _{\mbox{\scriptsize max}}$ and the red continuous line is $j _{z} (\theta) / j _{\mbox{\scriptsize max}}$.  The right panel shows the axial current components: the blue dashed  line corresponds to $j _{5y} (\theta) / j _{5\mbox{\scriptsize max}}$ and the red continuous line is $j _{5z} (\theta) / j _{5\mbox{\scriptsize max}}$. The effects of pseudo-fields can be increased according to the strain. In the numerical calculations we have used $B _{5}=0.5$ T; however, strain-induced pseudo-magnetic fields (on the order of 3 Tesla) were recently observed in strained crystals of Re-doped MoTe$_{2}$ \cite{PhysRevB.100.115105}. 

The analysis of this subsection can be extended to situations involving more deformation profiles. In particular, it would be interesting to analyse the effect of an axial pseudo-electric field, which unlike the deformation potential considered above,  couple with opposite sign to the nodes of opposite chirality.

\subsection{Rotational strain}

\begin{figure}
    \centering
    \includegraphics[width=0.3\textwidth]{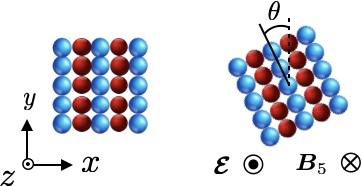}
    \caption{Schematic representation of a cross-section of the wire-shaped nanowire twisted by an angle $\theta$. } \label{Fig_Twisted_WSM}
\end{figure}

In three dimensions, the antisymmetric part of the strain tensor, $\omega _{ij} = \tfrac{1}{2} \left(  \partial u _{j} / \partial x ^{i}  - \partial u _{i} / \partial x ^{j} \right)$,  also gives rise to  pseudo-fields in Weyl semimetals. Physically, it is related to infinitesimal rotations by a vector $\boldsymbol{\Omega}$ with the components given by $\Omega _{k} = \tfrac{1}{2} \epsilon _{ijk} \omega _{ij}$. One can further see that this vector is related to the deformation vector $\boldsymbol{u}$ by 
\begin{align}
\boldsymbol{\Omega} (\boldsymbol{r}) = \frac{1}{2} \, \nabla \times \boldsymbol{u} (\boldsymbol{r}) . \label{Vorticity}
\end{align}
A full discussion on the effects of rotational strain in Dirac matter is presented in Ref. \cite{PhysRevB.97.201404}. There, the authors derive the low-energy effective Hamiltonians for the electron-strain interactions around Weyl nodes associated with the antisymmetric part of the strain tensor. A distinguishing physical example, which we consider here, is a wire-shaped Weyl semimetal of length $L$ with an axis along the $z$-direction and the nodes separated by a distance $b$ along the axis. The displacement vector $\boldsymbol{u}$ that derives from twisting the sample an angle $\theta$ is given by:
\begin{align}
\boldsymbol{u} (\boldsymbol{r}) = \theta \frac{z}{L} (\boldsymbol{r} \times \hat{\boldsymbol{e}} _{z}) ,  \label{Deformation}
\end{align}
where $\boldsymbol{r}$ is the position relative to the origin located on the axis of the wire.  In Fig. \ref{Fig_Twisted_WSM} we show the effect of the deformation (\ref{Deformation}). The strain tensor associated with the deformation (\ref{Deformation}) is traceless, and therefore the deformation potential generated from $u _{ij}$ is zero.  The corresponding axial vector potential is given by $\boldsymbol{A} _{5} (\boldsymbol{r}) = \theta \frac{b}{2L} (y \hat{\boldsymbol{e}} _{x} - x \hat{\boldsymbol{e}} _{y})$, producing the uniform pseudo-magnetic field $\boldsymbol{B} _{5} = - \theta \frac{b}{L}   \hat{\boldsymbol{e}} _{z}$.  According to Eq. (\ref{Vorticity}), the rotation vector becomes $\boldsymbol{\Omega} (\boldsymbol{r}) = \frac{\theta}{2L} (\boldsymbol{r} -3 z \hat{\boldsymbol{e}} _{z}) $. This produces a deformation potential $\tilde{\Phi} (\boldsymbol{r}) = \boldsymbol{b} \cdot \boldsymbol{\Omega} (\boldsymbol{r}) = - \frac{\theta}{L} bz$, which is non-axial. The corresponding pseudo-electric field is $\boldsymbol{\mathcal{E}} = - \nabla \boldsymbol{\Omega} (\boldsymbol{r}) = \frac{\theta}{L} b \hat{\boldsymbol{e}} _{z}$, which is antiparallel to the pseudo-magnetic field $\boldsymbol{B} _{5}$, as depicted in Fig. \ref{Fig_Twisted_WSM}. It is interesting that these pseudo-fields induce a longitudinal current along the direction of the pseudo-fields:
\begin{align}
J _{z} = \sigma _{0} \left( 1 + \frac{B _{5} ^{2}}{5 B _{0} ^{2}} \right) \mathcal{E} . 
\end{align}
Note that the direction of the torsion-induced current can be controlled by the direction of the rotation: it is positive (negative) when the twisting is clockwise (anticlockwise). This configuration, supplemented by an external magnetic field, could also be used to test the covariant anomaly equations.

\section{Discussion} \label{conclusion}

In recent years,  the study of strain-induced transport has attracted great attention due to its potential applications for the development of straintronic devices. The most salient example is perhaps graphene, in which strain induces pseudo-fields which couples to the Dirac fermions oppositely in the two valleys. Other well-known examples are carbon nanotubes \cite{PhysRevLett.78.1932}, bilayer graphene \cite{PhysRevB.86.165448, PhysRevB.88.035404} and transition metal dichalcogenides \cite{PhysRevLett.113.077201, PhysRevB.92.195402}. More recently, the study of strain-induced pesudo-fields in Dirac materials has garnered a lot of attention since it could open a pathway to the development of the strain-induced chiralitytronics. In Weyl semimetals, for example,  these chirality selective pseudo-fields has lead to interesting strain-induced phenomena,  such as quantum oscillations \cite{PhysRevB.95.041201}, the chiral magnetic effect and the negative resistivity \cite{PhysRevB.94.241405, PhysRevX.6.041046}, the chiral torsional effect \cite{PhysRevLett.116.166601}, the acoustogalvanic effect \cite{PhysRevLett.124.126602}, among others.

In this paper we used the chiral kinetic theory approach, {at a finite temperature,} to investigate nonlinear transport phenomena in Weyl semimetals induced by electromagnetic fields and strain-induced pseudo-fields. Our main focus was the study of the planar Hall effect, the appearance of an in-plane transverse voltage in the presence of coplanar electromagnetic fields. Using the relaxation time approximation for the collision integral, we first derived general expressions for the nonlinear conductivity tensor and clearly differentiate the contributions arising from the Berry curvature and the orbital magnetic moment of charge carries, since the latter has been continuously disregarded in the analysis of the PHE, and we showed that it is as important as the Berry curvature.  Next, using the simplest linearly dispersing model for a Weyl semimetal with two nodes of opposite chiralities separated in momentum and energy, we obtained analytical expressions for the contributions to the magnetoconductivity tensor {at a finite temperature. A rough estimation of the finite temperature effect reveals that, at low temperatures ($T \sim 2$K, which is the temperature used to detect the PHE in topological insulators), the amplitude of the currents differ by a factor of $10^{-4}$ with respect to the result at $T=0$.} Our numerical calculations, presented in Figs. \ref{Conduc_Chiralities}-\ref{Conduc_3D}, reveal that the OMM contribution is more pronounced in the longitudinal magentoconductivity than in the planar Hall conductivity. Our general expressions for the total $\boldsymbol{J}$ and axial $\boldsymbol{J} _{5}$ currents,  (\ref{Total_Current}) and (\ref{Axial_Current}) respectively, include the effects of usual electromagnetic fields $\boldsymbol{E}$ and $\boldsymbol{B}$, besides the contributions arising from the pseudo-electric $\boldsymbol{E} _{5}$ and the pseudo-magnetic $\boldsymbol{B} _{5}$ fields. {In the absence of the axial electric field the longitudinal and planar Hall currents exhibit periodicity in the angle $\theta$, defined by the magnetic field as $\cos \theta = \boldsymbol{\hat{B}} _{5} \cdot \boldsymbol{\hat{E}}$; however, when the axial electric field is turned on, such periodicity goes down and the position of the extrema in the conductivities change.} Interestingly, these expressions show the possibility of inducing nonlinear transport phenomena by using only strain-induced pseudo-fields without needing real electromagnetic fields, which opens up new avenues for manipulating and controlling this effect.  Finally, we apply our results to two strained configurations with experimental possibilities: (i) bending a WSM thin film into a circular arc and (ii) applying torsion to a WSM wire. These configurations could also be useful in the investigation of the covariant anomaly equations (\ref{Chiral_Anomaly}) and (\ref{Charge_Anomaly}) of Weyl fermions.

The planar Hall effect in multilayer structures of NiFe/IrMn and NiFe/Cu/NiFe has been widely used in the development of high-performance magnetic field sensors, including those based on spin valves, giant magnetoresistance and tunneling magnetoresistance \cite{LUONG2018399, 8466005, QUYNH201698}. On the other hand, in Dirac materials, for example, recent magneto-transport measurements on single crystals of the magnetic Weyl semimetal Co$_3$Sn$_2$S$_2$ reported the observation of the PHE \cite{SHAMA2020166547}.   Along these lines, our results create opportunities for the design of novel devices or sensors that exploit pseudo-fields to detect the PHE in Weyl semimetals.  Finally, we note that the study of strain-induced PHE can be extended straightforwardly to include dynamical local deformations of the crystal, giving rise to an acoustic-induced PHE, thus enlarging the possibilities for the design of devices for chiralitytronics.

\acknowledgements{ L.M.O. was supported by the CONACyT PhD fellowship No. 834773. A.M.-R. has been partially supported by DGAPA-UNAM Project No. IA102722 and by Project CONACyT (M\'{e}xico) No. 428214.  {We are also indebted to the reviewers for their valuable comments and suggestions to improve the quality of the paper.} }
 
\

\appendix

\section{Derivation of the orbital magnetic moment contribution} \label{App_OMM}

In this Section we derive the formula for the orbital magnetic moment contribution $\sigma ^{(m , \alpha)} _{ij}$ to the magnetoconductivity tensor $\sigma ^{(\alpha)}  _{ij}$. To this end, we start with the general expression for the magnetoconductivity tensor, given by Eq. (\ref{Conductivity_Tensor}) and expand the factor  $D _{\alpha } ({\boldsymbol{k}} )$ and the equilibrium distribution function $f _{\alpha} ^{\mbox{\scriptsize eq}} ( \mathcal{E} _{\alpha} )$ in powers of the pseudo-magnetic field. Keeping terms up to quadratic order in the pseudo-magnetic field we obtain
\begin{align}
D _{\alpha } ({\boldsymbol{k}} ) & \approx 1 - \frac{e}{\hbar} \left[ {\boldsymbol{B}} _{\chi} \cdot {\boldsymbol{\Omega}} _{\alpha} ({\boldsymbol{k}}) \right] + \frac{e ^{2}}{\hbar ^{2}} \left[ {\boldsymbol{B}} _{\chi} \cdot {\boldsymbol{\Omega}} _{\alpha} ({\boldsymbol{k}}) \right] ^{2}   \notag \\[5pt] & \equiv 1 + D _{\alpha } ^{(1)} ({\boldsymbol{k}} )  + D _{\alpha } ^{(2)} ({\boldsymbol{k}} )   \label{Exp_D}
\end{align}
and
\begin{align}
f _{\alpha} ^{\mbox{\scriptsize eq}} ( \mathcal{E} _{\alpha}   ) \approx f _{\alpha} ^{\mbox{\scriptsize eq}} ( \mathcal{E} _{\alpha} ^{(0)}  ) +  \mathcal{E} _{\alpha} ^{(m)} f _{\alpha} ^{\mbox{\scriptsize eq} \, \prime } (\mathcal{E} _{\alpha} ^{(0)})+ \frac{1}{2} \mathcal{E} _{\alpha} ^{(m) 2 } f _{\alpha} ^{ \mbox{\scriptsize eq} \, \prime \prime } ( \mathcal{E} _{\alpha} ^{(0)})  ,  \label{Exp_f}
\end{align}
where $f _{\alpha} ^{\mbox{\scriptsize eq} \, \prime } (\mathcal{E} _{\alpha} ^{(0)}) = \frac{\partial  f _{\alpha} ^{\mbox{\scriptsize eq}} (\mathcal{E} _{\alpha} ^{(0)}) }{ \partial \mathcal{E} _{\alpha} ^{(0)}  } $  and $f _{\alpha} ^{ \mbox{\scriptsize eq} \, \prime \prime } ( \mathcal{E} _{\alpha} ^{(0)}) = \frac{\partial ^{2}  f _{\alpha} ^{\mbox{\scriptsize eq}} (\mathcal{E} _{\alpha} ^{(0)}) }{ \partial \mathcal{E} _{\alpha} ^{(0)2}  }  $.  Now we substitute these expansions into Eq. (\ref{Conductivity_Tensor}) and separate the velocity as ${\boldsymbol{v}} _{\alpha } = {\boldsymbol{v}} _{\alpha } ^{(0)} + {\boldsymbol{v}} _{\alpha } ^{(m)} $.  Keeping the terms quadratic in the pseudo-magnetic field we obtain:
\begin{widetext}
\begin{align}
\sigma ^{(m , \alpha)} _{ij} &= - e ^{2} \tau \int \frac{d ^{3} {\boldsymbol{k}}}{(2 \pi ) ^{3} } \Bigg\{\ \!\!  \left[ v _{\alpha i} ^{(m)} v _{\alpha j} ^{(m)}   f _{\alpha} ^{\mbox{\scriptsize eq} \, \prime } (\mathcal{E} _{\alpha} ^{(0)})  +  (v _{\alpha i} ^{(0)} v _{\alpha j} ^{(m)}  + v _{\alpha i} ^{(m)} v _{\alpha j} ^{(0)} ) \mathcal{E} _{\alpha} ^{(m)}  f _{\alpha} ^{\mbox{\scriptsize eq} \, \prime \prime } (\mathcal{E} _{\alpha} ^{(0)})  + \frac{1}{2} v _{\alpha i} ^{(0)} v _{\alpha j} ^{(0)} \mathcal{E} _{\alpha} ^{(m) 2 }   f _{\alpha} ^{\mbox{\scriptsize eq} \, \prime \prime \prime } (\mathcal{E} _{\alpha} ^{(0)})   \right]  \notag \\[5pt] & \phantom{=} +  \frac{e}{\hbar}  \left[ \Omega _{\alpha k}  \left( B _{\chi i} (v _{\alpha j} ^{(0)} v _{\alpha k} ^{(m)}  + v _{\alpha j} ^{(m)} v _{\alpha k} ^{(0)} ) + B _{\chi j} (v _{\alpha i} ^{(0)} v _{\alpha k} ^{(m)}  + v _{\alpha i} ^{(m)} v _{\alpha k} ^{(0)} ) -  B _{\chi k} (v _{\alpha i} ^{(0)} v _{\alpha j} ^{(m)}  + v _{\alpha i} ^{(m)} v _{\alpha j} ^{(0)} ) \right)   f _{\alpha} ^{\mbox{\scriptsize eq} \, \prime } (\mathcal{E} _{\alpha} ^{(0)}) \right. \notag \\[5pt] & \phantom{=}   \left.   + \; \Omega _{\alpha k }  \left( B _{\chi i} v _{\alpha j} ^{(0)} v _{\alpha k} ^{(0)} + B _{\chi j} v _{\alpha i} ^{(0)} v _{\alpha k} ^{(0)} -  B _{\chi k}  v _{\alpha i} ^{(0)} v _{\alpha j} ^{(0)} \right)    \mathcal{E} _{\alpha} ^{(m)} f _{\alpha} ^{\mbox{\scriptsize eq} \, \prime \prime } (\mathcal{E} _{\alpha} ^{(0)})    \right]  \, \, \Bigg\}\  ,  \label{omm1}
\end{align}
where we have subtracted the field-independent and Berry curvature contributions, given by equations (\ref{Conductivity_Tensor_0B}) and (\ref{Conductivity_Tensor_Berry}) respectively. To simplify this expression, on the one hand, we observe that: 
\begin{align}
\partial _{k _i} \partial _{k _j} \left( \frac{1}{2 \hbar ^{2}}  \mathcal{E} _{\alpha} ^{(m) 2 } f _{\alpha} ^{\mbox{\scriptsize eq} \, \prime } (\mathcal{E} _{\alpha} ^{(0)}) \right) &=  \frac{1}{\hbar}  \left( \partial _{k _j} v _{\alpha i } ^{(m)} \right) \mathcal{E} _{\alpha} ^{(m)  }  f _{\alpha} ^{\mbox{\scriptsize eq} \, \prime } (\mathcal{E} _{\alpha} ^{(0)}) +    \frac{1}{2 \hbar } \left( \partial _{k _j} v _{\alpha i } ^{(0)} \right) \mathcal{E} _{\alpha} ^{(m) 2 }  f _{\alpha} ^{\mbox{\scriptsize eq} \, \prime \prime } (\mathcal{E} _{\alpha} ^{(0)}) +   \left[ v _{\alpha i} ^{(m)} v _{\alpha j} ^{(m)}   f _{\alpha} ^{\mbox{\scriptsize eq} \, \prime } (\mathcal{E} _{\alpha} ^{(0)}) \right. \notag \\[5pt] & \phantom{=}  \left. \;  + \;  (v _{\alpha i} ^{(0)} v _{\alpha j} ^{(m)}  + v _{\alpha i} ^{(m)} v _{\alpha j} ^{(0)} ) \mathcal{E} _{\alpha} ^{(m)}  f _{\alpha} ^{\mbox{\scriptsize eq} \, \prime \prime } (\mathcal{E} _{\alpha} ^{(0)})  + \frac{1}{2} v _{\alpha i} ^{(0)} v _{\alpha j} ^{(0)} \mathcal{E} _{\alpha} ^{(m) 2 }   f _{\alpha} ^{\mbox{\scriptsize eq} \, \prime \prime \prime } (\mathcal{E} _{\alpha} ^{(0)})   \right] , \label{Aux1-omm}
\end{align}
where we have used that $\nabla _{ \boldsymbol{k} } f _{\alpha} ^{\mbox{\scriptsize eq} } (\mathcal{E} _{\alpha} ^{(0)}) = \hbar  \boldsymbol{v} _{\alpha} ^{(0)}  f _{\alpha} ^{\mbox{\scriptsize eq} \, \prime } (\mathcal{E} _{\alpha} ^{(0)}) $. The term in square brackets is exactly the same that appears in the first line of Eq. (\ref{omm1}).  Besides, upon integration of Eq. (\ref{Aux1-omm}) over the Brillouin zone, the left-hand side vanishes due to the periodic boundary conditions. Hence, the first line of Eq. (\ref{omm1}) can be replaced by the two first terms in the right-hand side of Eq. (\ref{Aux1-omm}).  On the other hand, the second term in square brakets in Eq. (\ref{omm1}) can be further simplified in terms of the vector $\boldsymbol{Q} _{\alpha} \equiv  \boldsymbol{\Omega} _{\alpha} \times (  \boldsymbol{v} ^{(0)} _{\alpha} \times  \boldsymbol{B}  _{\chi } )$. We obtain
\begin{align}
\sigma ^{(m , \alpha)} _{ij} &= - e ^{2} \tau \int \frac{d ^{3} {\boldsymbol{k}}}{(2 \pi ) ^{3} } \Bigg\{\ \!\! - \left[ \frac{1}{\hbar}  \left( \partial _{k _j} v _{\alpha i } ^{(m)} \right) \mathcal{E} _{\alpha} ^{(m)  }  f _{\alpha} ^{\mbox{\scriptsize eq} \, \prime } (\mathcal{E} _{\alpha} ^{(0)}) +    \frac{1}{2 \hbar } \left( \partial _{k _j} v _{\alpha i } ^{(0)} \right) \mathcal{E} _{\alpha} ^{(m) 2 }  f _{\alpha} ^{\mbox{\scriptsize eq} \, \prime \prime } (\mathcal{E} _{\alpha} ^{(0)})    \right]  \notag \\[5pt] & \phantom{=} +  \frac{e}{\hbar}    \left[  - \left( {Q} _{\alpha i} v _{\alpha j} ^{(m)} + {Q} _{\alpha j} v _{\alpha i} ^{(m)} \right) f _{\alpha} ^{\mbox{\scriptsize eq} \, \prime  } (\mathcal{E} _{\alpha} ^{(0)})    +  \left( B _{\chi i}   v _{\alpha j} ^{(0)}  + B _{\chi j}   v _{\alpha i} ^{(0)} \right) \boldsymbol{\Omega} _{\alpha} \cdot  \left(   \boldsymbol{v} ^{(m)} _{\alpha}  f _{\alpha} ^{\mbox{\scriptsize eq} \, \prime   } (\mathcal{E} _{\alpha} ^{(0)})   +    \boldsymbol{v} ^{(0)} _{\alpha}   \mathcal{E} _{\alpha} ^{(m)} f _{\alpha} ^{\mbox{\scriptsize eq} \, \prime \prime } (\mathcal{E} _{\alpha} ^{(0)})   \right) \right] \notag \\[5pt] & \phantom{=}  -  \frac{e}{\hbar}  ( \boldsymbol{\Omega} _{\alpha} \cdot  \boldsymbol{B} _{\chi} )  v _{\alpha i} ^{(0)} v _{\alpha j} ^{(0)}    \mathcal{E} _{\alpha} ^{(m)} f _{\alpha} ^{\mbox{\scriptsize eq} \, \prime \prime } (\mathcal{E} _{\alpha} ^{(0)})    \, \, \Bigg\}\  .  \label{omm2}
\end{align}
Now, using the fact that $\frac{1}{\hbar} \nabla _{ \boldsymbol{k} } \left[  \mathcal{E} _{\alpha} ^{(m)}  \,  f _{\alpha} ^{\mbox{\scriptsize eq} \, \prime } (\mathcal{E} _{\alpha} ^{(0)})  \right] =   \boldsymbol{v} ^{(m)} _{\alpha}  f _{\alpha} ^{\mbox{\scriptsize eq} \, \prime   } (\mathcal{E} _{\alpha} ^{(0)})   +    \boldsymbol{v} ^{(0)} _{\alpha}   \mathcal{E} _{\alpha} ^{(m)} f _{\alpha} ^{\mbox{\scriptsize eq} \, \prime \prime } (\mathcal{E} _{\alpha} ^{(0)}) $ and integrating by parts we get
\begin{align}
\sigma ^{(m , \alpha)} _{ij} &=   \frac{e ^{3} \tau}{\hbar}   \int \frac{d ^{3} {\boldsymbol{k}}}{(2 \pi ) ^{3} } \Bigg\{\ \!\!  \left( {Q} _{\alpha i} v _{\alpha j} ^{(m)} + {Q} _{\alpha j} v _{\alpha i} ^{(m)} \right) f _{\alpha} ^{\mbox{\scriptsize eq} \, \prime  } (\mathcal{E} _{\alpha} ^{(0)})  +   \mathcal{E} _{\alpha} ^{(m)}  \left[      ( \boldsymbol{\Omega} _{\alpha} \cdot  \boldsymbol{B} _{\chi} )  v _{\alpha i} ^{(0)} v _{\alpha j} ^{(0)}  + \frac{1}{2e} \mathcal{E} _{\alpha} ^{(m) }   \left( \partial _{k _j} v _{\alpha i } ^{(0)} \right)  \right] f _{\alpha} ^{\mbox{\scriptsize eq} \, \prime \prime } (\mathcal{E} _{\alpha} ^{(0)})   \notag \\[5pt] & \phantom{=} +   \frac{1}{e} \mathcal{E} _{\alpha} ^{(m)  }   \left[  \left( \partial _{k _j} v _{\alpha i } ^{(m)} \right) +  \frac{e}{\hbar}   \nabla _{ \boldsymbol{k} }  \cdot \left[  \boldsymbol{\Omega} _{\alpha} ( B _{\chi i}   v _{\alpha j} ^{(0)}  + B _{\chi j}   v _{\alpha i} ^{(0)} )   \right] \right]   f _{\alpha} ^{\mbox{\scriptsize eq} \, \prime  } (\mathcal{E} _{\alpha} ^{(0)})    \, \, \Bigg\}\  , \label{omm3}
\end{align}
where we have used the periodicity of the Brillouin zone.  Finally,  using the symmetry of the tensor under the interchange $i \leftrightarrow j$ and the expression $\mathcal{E} _{\alpha} ^{(m)}  = -  {\boldsymbol{m}} _{\alpha}   \cdot {\boldsymbol{B}} _{\chi}$ we obtain
\begin{align}
\sigma ^{(m , \alpha)} _{ij} &=   \frac{e ^{3} \tau}{\hbar}   \int \frac{d ^{3} {\boldsymbol{k}}}{(2 \pi ) ^{3} } \Bigg\{\ \!\!   {Q} _{\alpha i} v _{\alpha j} ^{(m)}  f _{\alpha} ^{\mbox{\scriptsize eq} \, \prime  } (\mathcal{E} _{\alpha} ^{(0)})  +  \frac{1}{2} \mathcal{E} _{\alpha} ^{(m)}   {\boldsymbol{B}} _{\chi}  \cdot \left[    \boldsymbol{\Omega} _{\alpha}    v _{\alpha i} ^{(0)} v _{\alpha j} ^{(0)}  -  {\boldsymbol{m}} _{\alpha}  \frac{1}{2e}   \left( \partial _{k _j} v _{\alpha i } ^{(0)} \right)  \right] f _{\alpha} ^{\mbox{\scriptsize eq} \, \prime \prime } (\mathcal{E} _{\alpha} ^{(0)})   \notag \\ & \phantom{=} +   \frac{1}{e} \, \mathcal{E} _{\alpha} ^{(m)  }   \left[  \frac{e}{\hbar}   \nabla _{ \boldsymbol{k} }  \cdot (  \boldsymbol{\Omega} _{\alpha} B _{\chi i}   v _{\alpha j} ^{(0)}  )  +  \frac{1}{2}  \left( \partial _{k _j} v _{\alpha i } ^{(m)} \right) \right]   f _{\alpha} ^{\mbox{\scriptsize eq} \, \prime  } (\mathcal{E} _{\alpha} ^{(0)})    \, \, \Bigg\}\  +  (i \leftrightarrow j) . \label{omm4}
\end{align}
This result suggests the definition of the tensors in Eq. (\ref{Tensors}) and yields the final expression for the orbital magnetic moment contribution, given by Eq. (\ref{Conductivity_Tensor_OMM}).

\end{widetext}

\section{Computation of the conductivity tensors} \label{Det_Calc}

Here we evaluate in detail the different contributions to the magnetoconductivity tensor defined in Section \ref{Kinetic} for a Weyl semimetal. Owing to the rotational symmetry of the problem, the required integrals can be performed straightforwardly by using  the spherical coordinate system ${\boldsymbol{k}} = k (\sin \theta \cos \phi \hat{{\boldsymbol{e}}}_{x} + \sin \theta \sin \phi \hat{{\boldsymbol{e}}}_{y} + \cos \theta \hat{{\boldsymbol{e}}}_{z} ) $ and the volume element $d ^{3} {\boldsymbol{k}} = k^{2} dk d \Omega $, being $d \Omega $ the differential solid angle.  Note that the difference between the chemical potential and the energy shift of the node determines the band index, i.e. $s = \mbox{sgn} (\mu - b _{0 \chi})$, as evinced in figure \ref{Fig_Cones}. This is so since $\mu > b _{0 \chi}$ ($\mu < b _{0 \chi}$) implies $s = 1$ ($s=-1$). Therefore we take $\mu _{\chi} = s \mu _{0 \chi}$, with $\mu _{0 \chi} = \vert \mu - b _{0 \chi} \vert > 0$.  {Besides, we work at finite temperature. }


Let us first consider the $B$-independent conductivity, given by Eq. (\ref{Conductivity_Tensor_0B}). Substituting the required components of the band velocity ${\boldsymbol{v}} _{s} ^{(0)} = s v _{F} \hat{{\boldsymbol{k}}}$ one gets
\begin{align}
\sigma ^{(0, \alpha)}  _{ij} (T)  = - \frac{e ^{2} v _{F} ^{2} \tau}{8 \pi ^{3}} \int d ^{3} {\boldsymbol{k}} \,  \hat{k} _{i} \hat{k} _{j}  \,  \frac{\partial  f _{\alpha} ^{\mbox{\scriptsize eq}} (\mathcal{E} _{\alpha} ^{(0)}) }{ \partial \mathcal{E} _{\alpha} ^{(0)}  }  .
\end{align}
Owing to the rotational symmetry, the angular integration becomes $\int d \Omega \, \hat{k} _{i} \hat{k} _{j} = \int d \Omega \, \frac{1}{3} \delta _{ij} = \frac{4 \pi }{3}  \delta _{ij}$. Therefore we obtain
\begin{align}
\sigma ^{(0, \alpha )}  _{ij} (T)  =   \frac{e ^{2} v _{F} ^{2} \tau}{6 \pi ^{2}} \delta _{ij} \int _{0} ^{\infty} dk \, k ^{2}  \,  \frac{1}{k _{B} T}  \frac{e ^{ \frac{ \mathcal{E} _{\alpha} ^{(0)} - \mu _{\chi} }{k _{B} T} }}{ \left( 1 + e ^{ \frac{ \mathcal{E} _{\alpha} ^{(0)} - \mu _{\chi} }{k _{B} T} } \right) ^{2} }  ,
\end{align}
where we have used the Fermi-Dirac distribution. Using the fact that $\mathcal{E} ^{(0)} _{\alpha} ({\boldsymbol{k}}) = b _{0 \chi} + s \hbar v _{F} k$ and $\mu _{\chi} = s \mu _{0 \chi}$, with $\mu _{0 \chi} = \vert \mu - b _{0 \chi} \vert > 0$, this result can be written in the simple form  
\begin{align}
 \sigma _{ij} ^{(\chi , 0)} (T) =   \frac{e ^{2} \tau \mu _{0 \chi } ^{2}  }{ 6 \pi ^{2} \hbar ^{3} v _{F} } \delta _{ij}  \,  f _{2} (\Lambda _{\chi} )  ,  \label{B0_Conductivity_Temp}
\end{align}
with $\Lambda _{\chi} \equiv k _{B} T / \mu _{0 \chi }$ and where we have introduced the function
\begin{align}
{f _{n} (y) \equiv  \frac{1}{y} \int _{0} ^{\infty} dx \, x ^{n} \,   \frac{e ^{ (x - 1 )/ y }}{ \left[ 1 + e ^{ (x - 1 )/ y  } \right] ^{2} }  .  } \label{F_function}
\end{align}
It is interesting that $\lim _{y \to 0} f _{n} (y) = 1$, which applied to our result of Eq. (\ref{B0_Conductivity_Temp}) corresponds to the zero temperature limit. 
 
We now turn to the Berry curvature contribution, given by Eq. (\ref{Conductivity_Tensor_Berry}).  Using the Berry curvature ${\boldsymbol{\Omega}} _{\alpha}$ and band velocity $\boldsymbol{v} _{\alpha} ^{(0)}$, given in Eq. (\ref{B_Curvature}), one finds ${\boldsymbol{Q}} _{\alpha}  =  - \chi  \frac{v _{F} }{2k ^{2}}  \hat{{\boldsymbol{k}}} \times (  \hat{{\boldsymbol{k}}} \times {\boldsymbol{B}} _{\chi}   )$.  Therefore, Eq. (\ref{Conductivity_Tensor_Berry}) yields
\begin{align}
\sigma _{ij} ^{(\alpha ,  \Omega )} (\boldsymbol{B} _{\chi} , T ) &= -  \frac{e ^{4}  v _{F} ^{2} \tau }{32 \pi ^{3} \hbar ^{2} }  \int d ^{3} {\boldsymbol{k}} \,  \frac{1}{k ^{4}} \left[ \hat{{\boldsymbol{k}}} \times (  \hat{{\boldsymbol{k}}} \times {\boldsymbol{B}} _{\chi} )  \right] _{i} \notag \\[5pt] & \phantom{=} \times \left[ \hat{{\boldsymbol{k}}} \times ( \hat{{\boldsymbol{k}}} \times {\boldsymbol{B}} _{\chi} ) \right] _{j}  \,   \frac{\partial  f _{\alpha} ^{\mbox{\scriptsize eq}} (\mathcal{E} _{\alpha} ^{(0)}) }{ \partial \mathcal{E} _{\alpha} ^{(0)}  }  .  \label{Int_Berry}
\end{align}
Algebraic manipulation of the integrand produces 
\begin{align}
\sigma _{ij} ^{(\alpha ,  \Omega )} (\boldsymbol{B} _{\chi} , T) &= - \frac{e ^{4}  v _{F} ^{2} \tau }{32 \pi ^{3} \hbar ^{2} } \!  \int \!  \frac{d k }{k ^{2}} \,  \frac{\partial  f _{\alpha} ^{\mbox{\scriptsize eq}} (\mathcal{E} _{\alpha} ^{(0)}) }{ \partial \mathcal{E} _{\alpha} ^{(0)}  }   \int   d \Omega \Big[   B _{\chi i }  B _{\chi j }   \notag \\[5pt]  & \hspace{-0.3cm} - ( \hat{{\boldsymbol{k}}} \cdot {\boldsymbol{B}} _{\chi} ) ( \hat{k} _{i}  B _{\chi j }  + \hat{k} _{j} B _{\chi i } ) + ( \hat{{\boldsymbol{k}}} \cdot {\boldsymbol{B}} _{\chi} ) ^{2} \hat{k} _{i} \hat{k} _{j}   \Big] . \label{Int_Berry2}
\end{align}
These integrals are simple, but not straightforward. The necessary angular integral over products of rectangular components of $\hat{{\boldsymbol{k}}}$ is readily found to be
\begin{align}
\int d \Omega \,  \hat{k} _{i} \hat{k} _{j} \hat{k} _{l} \hat{k} _{m}  = \frac{4 \pi}{15}  \left( \delta _{ij} \delta _{lm} + \delta _{il} \delta _{jm} + \delta _{im} \delta _{jl} \right) . \label{Ang_Integral}
\end{align}
On the other hand, the required radial integral can be expressed in terms of the function $f _{n} (\Lambda _{\chi})$, defined by Eq. (\ref{F_function}), with $n=-2$, i.e.
\begin{align}
\int _{0} ^{\infty} dk \, \frac{1}{k ^{2}}  \, \frac{\partial f _{\alpha} ^{\mbox{\scriptsize eq}} (\mathcal{E} _{\alpha} ^{(0)})  }{ \partial \mathcal{E} _{\alpha} ^{(0)} } &=   - \frac{\hbar v _{F} }{\mu _{0 \chi } ^{2} }  f _{ - 2} (\Lambda _{\chi}) .  \label{Rad_Integral}
\end{align}
Substituting these results into Eq. (\ref{Int_Berry2}) we establish Eq. (\ref{Conductivity_Tensor_Berry_Final}).

We now evaluate the orbital magnetic moment contribution $\sigma _{ij} ^{(\chi , m )} ({\boldsymbol{B}} _{\chi},T)$, given by Eq. (\ref{Conductivity_Tensor_OMM}).  To this end,  we require the corrections to the energy $ \mathcal{E} _{\alpha} ^{(m)}$ and band velocity ${\boldsymbol{v}} _{\alpha } ^{(m)} = \frac{1}{\hbar} \nabla _{{\boldsymbol{k}}} \mathcal{E} _{\alpha} ^{(m)} $ due to the orbital magnetic moment. Using the OMM ${\boldsymbol{m}} _{\alpha}$, given in Eq. (\ref{B_Curvature}), one finds $ \mathcal{E} _{\alpha} ^{(m)} = \frac{\chi e v_{F}}{2 k} \hat{{\boldsymbol{k}}} \cdot {\boldsymbol{B}} _{\chi}$, and
\begin{align}
{\boldsymbol{v}} _{\alpha } ^{(m)} ({\boldsymbol{k}}) =  \frac{ \chi e v _{F}  }{2 \hbar   }  \frac{ {\boldsymbol{B}} _{\chi} - 2  \hat{{\boldsymbol{k}}} ( \hat{{\boldsymbol{k}}} \cdot {\boldsymbol{B}} _{\chi} )}{k ^{2}}   .  \label{OMM_velocity}
\end{align}
Now we compute separately the three integrals in Eq. (\ref{Conductivity_Tensor_OMM}) in the order they appear, i.e.  $\sigma _{ij} ^{(\chi , m )} = \sigma _{1ij} ^{(\chi , m )} + \sigma _{2ij} ^{(\chi , m )} + \sigma _{3ij} ^{(\chi , m )} $. Let us consider first:
\begin{align}
\sigma _{1 ij} ^{(\chi , m )} ({\boldsymbol{B}} _{\chi},T) =   \frac{2 e ^{3} \tau}{\hbar} \! \int \! \frac{d ^{3} {\boldsymbol{k}} }{(2 \pi ) ^{3}}   {Q} _{\alpha i} v _{\alpha j} ^{(m)}    \frac{\partial f _{\alpha} ^{\mbox{\scriptsize eq}} (\mathcal{E} _{\alpha} ^{(0)}  ) }{ \partial \mathcal{E} _{\alpha} ^{(0)}  } .
\end{align}
Substituting the function ${Q} _{\alpha i}$ and the velocity (\ref{OMM_velocity}) we have
\begin{align}
\sigma _{1 ij} ^{(\chi , m )} ({\boldsymbol{B}} _{\chi},T) & =   -  \frac{  e ^{4} v _{F} ^{2} \tau}{16 \pi ^{3} \hbar ^{2} } \int  d ^{3} {\boldsymbol{k}} \, \frac{1}{k ^{4}} \, [  \hat{{\boldsymbol{k}}} \times (  \hat{{\boldsymbol{k}}} \times {\boldsymbol{B}} _{\chi} ) ] _{i}  \notag \\[5pt] & \phantom{=}  \times  [ {\boldsymbol{B}} _{\chi} - 2  \hat{{\boldsymbol{k}}} ( \hat{{\boldsymbol{k}}} \cdot {\boldsymbol{B}} _{\chi} ) ] _{j}  \,  \frac{\partial  f _{\alpha} ^{\mbox{\scriptsize eq}} (\mathcal{E} _{\alpha} ^{(0)}) }{ \partial \mathcal{E} _{\alpha} ^{(0)}  } . 
\end{align}
Manipulating the integrand we obtain
\begin{align}
\sigma _{1 ij} ^{(\chi , m )}  & =    \frac{  e ^{4} v _{F} ^{2} \tau}{16 \pi ^{3} \hbar ^{2} } \int   \frac{dk}{k ^{2}} \,  \frac{\partial  f _{\alpha} ^{\mbox{\scriptsize eq}} (\mathcal{E} _{\alpha} ^{(0)}) }{ \partial \mathcal{E} _{\alpha} ^{(0)}  }  \int d \Omega\,   \Big[   B _{\chi i }  B _{\chi j }  \notag \\[5pt] & \phantom{=}  + 2 \hat{k} _{i} \hat{k} _{j}  ( \hat{{\boldsymbol{k}}} \cdot {\boldsymbol{B}} _{\chi} ) ^{2} - ( \hat{{\boldsymbol{k}}} \cdot {\boldsymbol{B}} _{\chi} ) ( \hat{k} _{i} B _{\chi j } + 2  \hat{k} _{j} B _{\chi i} ) \Big] . 
\end{align}
The angular integration can be performed by using the formula (\ref{Ang_Integral}), and the radial integration is given by Eq. (\ref{Rad_Integral}). These results imply:
\begin{align}
\sigma _{1 ij} ^{(\chi , m )} ({\boldsymbol{B}} _{\chi},T) & =  -   \frac{  e ^{4} v _{F} ^{3} \tau}{30 \pi ^{2} \hbar   \mu _{0 \chi } ^{2} }   (\delta _{ij} B _{\chi} ^{2} + 2  B _{\chi i }  B _{\chi j }  )  f _{ - 2} (\Lambda _{\chi})    . \label{Sigma1_fin}
\end{align}
We now consider the second term  in Eq. (\ref{Conductivity_Tensor_OMM}), namely
\begin{align}
\sigma _{2 ij} ^{(\chi , m )} ({\boldsymbol{B}} _{\chi}) =  \frac{2 e ^{3} \tau}{\hbar} \! \int \! \frac{d ^{3} {\boldsymbol{k}} }{(2 \pi ) ^{3}} \,    \frac{1}{e} \mathcal{E} _{\alpha} ^{(m)}  \nabla _{{\boldsymbol{k}}} \cdot  \boldsymbol{T} _{\alpha ij}   \frac{\partial f _{\alpha} ^{\mbox{\scriptsize eq}} (\mathcal{E} _{\alpha} ^{(0)}  ) }{ \partial \mathcal{E} _{\alpha} ^{(0)}  }  , \label{Sigma2}
\end{align}
where $\boldsymbol{T} _{\alpha ij}$ is defined in Eq. (\ref{Tensors}). Using the Berry curvature ${\boldsymbol{\Omega}} _{\alpha}$ and the contributions to the band velocity $\boldsymbol{v} _{\alpha} ^{(0)}$ and $\boldsymbol{v} _{\alpha} ^{(m)}$, one finds
\begin{align}
\boldsymbol{T} _{\alpha ij} = \frac{\chi e v _{F} }{4 \hbar k ^{2} }   \left[ \hat{\boldsymbol{e}} _{i}     B _{\chi j } - 2 \hat{k} _{j} \left( B _{\chi i}  \hat{{\boldsymbol{k}}} + \hat{\boldsymbol{e}} _{i}    \, \hat{{\boldsymbol{k}}} \cdot {\boldsymbol{B}} _{\chi}  \right)  \right] ,
\end{align}
wherefrom we obtain
\begin{align}
\nabla _{{\boldsymbol{k}}} \cdot  \boldsymbol{T} _{\alpha ij}  &=   \frac{\chi e v _{F} }{2 \hbar k ^{3} }  \Big[ ( 4 \hat{k} _{i} \hat{k} _{j} - \delta _{ij}  )   \hat{\boldsymbol{k}} \cdot  {\boldsymbol{B}} _{\chi}  -  \hat{k} _{i} B _{\chi j } - \hat{k} _{j}  B _{\chi i }  \Big] . 
\end{align}
Inserting this result into Eq. (\ref{Sigma2}) and manipulating the integral we have
\begin{align}
\!\! \sigma _{2 ij} ^{(\chi , m )} \!  &= -  \frac{  e ^{4}  v _{F} ^{2}  \tau}{ 16 \pi ^{3} \hbar ^{2} }  \int  \frac{dk}{k ^{2}}  \frac{\partial f _{\alpha} ^{\mbox{\scriptsize eq}} (\mathcal{E} _{\alpha} ^{(0)}  ) }{ \partial \mathcal{E} _{\alpha} ^{(0)}  }    \, \int d \Omega \,  \hat{{\boldsymbol{k}}} \cdot {\boldsymbol{B}} _{\chi} \notag \\[5pt] & \phantom{=} \times \Big[ (  \delta _{ij}  - 4 \hat{k} _{i} \hat{k} _{j}  )   \hat{\boldsymbol{k}} \cdot  {\boldsymbol{B}} _{\chi}  +  \hat{k} _{i} B _{\chi j } + \hat{k} _{j}  B _{\chi i }  \Big]   . 
\end{align}
Finally, using the integrals (\ref{Ang_Integral}) and (\ref{Rad_Integral}) one gets
\begin{align}
\sigma _{2 ij} ^{(\chi , m )} ({\boldsymbol{B}} _{\chi},T) =  \frac{  e ^{4}  v _{F} ^{3}  \tau}{ 60 \pi ^{2} \hbar  \mu _{0 \chi } ^{2} }  (    \delta _{ij}  B _{\chi } ^{2} + 2  B _{\chi i } B _{\chi j }  ) f _{ - 2} (\Lambda _{\chi})  . \label{Sigma2_fin}
\end{align}
The last term we have to evaluate is:
\begin{align}
\sigma _{3ij} ^{(\chi , m )} ({\boldsymbol{B}} _{\chi},T) =  \frac{  e ^{3} \tau}{\hbar} \! \int \! \frac{d ^{3} {\boldsymbol{k}} }{(2 \pi ) ^{3}}      \mathcal{E} _{\alpha} ^{(m)}   \boldsymbol{B} _{\chi} \cdot \boldsymbol{V} _{\alpha ij }   \frac{\partial ^{2} f _{\alpha} ^{\mbox{\scriptsize eq}} (\mathcal{E} _{\alpha} ^{(0)}  ) }{ \partial \mathcal{E} _{\alpha} ^{(0)2}  } ,
\end{align}
where $\boldsymbol{V} _{\alpha ij }$ is defined in Eq. (\ref{Tensors}). Using the Berry curvature ${\boldsymbol{\Omega}} _{\alpha}$ and the orbital magnetic moment ${\boldsymbol{m}} _{\alpha}$, given by Eq. (\ref{B_Curvature}), we find
\begin{align}
\boldsymbol{V} _{\alpha ij} = s \chi  v _{F} ^{2}  \frac{\hat{{\boldsymbol{k}}}}{4k ^{2}}  ( \delta _{ij} - 3 \hat{k} _{i}  \hat{k} _{j} ), 
\end{align} 
and the integral to be solved is
\begin{align}
\sigma _{3ij} ^{(\chi , m )} ({\boldsymbol{B}} _{\chi},T) &=    \frac{  e ^{4} v _{F} ^{3}  \tau}{64 \pi ^{3}  \hbar} \int d ^{3} {\boldsymbol{k}} \,   \frac{s }{ k ^{3} }     ( \boldsymbol{B} _{\chi} \cdot \hat{{\boldsymbol{k}}} ) ^{2} ( \delta _{ij} - 3 \hat{k} _{i}  \hat{k} _{j} )  \notag \\[5pt] & \hspace{2.5cm} \times \frac{\partial ^{2} f _{\alpha} ^{\mbox{\scriptsize eq}} (\mathcal{E} _{\alpha} ^{(0)}  ) }{ \partial \mathcal{E} _{\alpha} ^{(0)2}  }  .
\end{align}
Further algebraic manipulations yield
\begin{align}
\sigma _{3ij} ^{(\chi , m )} ({\boldsymbol{B}} _{\chi},T) &=   \frac{  e ^{4} v _{F} ^{3}  \tau}{64 \pi ^{3}  \hbar} \int d \Omega \,   ( \boldsymbol{B} _{\chi} \cdot \hat{{\boldsymbol{k}}} ) ^{2} ( \delta _{ij} - 3 \hat{k} _{i}  \hat{k} _{j} )  \notag \\[5pt] & \hspace{0.7cm} \times \left[ s \int  \frac{dk}{k} \frac{\partial ^{2} f _{\alpha} ^{\mbox{\scriptsize eq}} (\mathcal{E} _{\alpha} ^{(0)}  ) }{ \partial \mathcal{E} _{\alpha} ^{(0)2}  } \right]      .
\end{align}
The angular integration is directly evaluated with the help of the result (\ref{Ang_Integral}). For the radial integration, starting from the function $f _{n} (y)$ given by Eq. (\ref{Rad_Integral}), one can spot the identity
 \begin{align}
s \int  \frac{dk}{k}   \frac{\partial ^{2} f _{\alpha} ^{\mbox{\scriptsize eq}} (\mathcal{E} _{\alpha} ^{(0)}  ) }{ \partial \mathcal{E} _{\alpha} ^{(0) 2}  }    &=  - \frac{1}{ \mu _{0 \chi } ^{2} }  f _{-2} ( \Lambda _{\chi} ) .
\end{align}
All in all,  the final result is
\begin{align}
\sigma _{3ij} ^{(\chi , m )} ({\boldsymbol{B}} _{\chi},T) &=   \frac{  e ^{4} v _{F} ^{3}  \tau}{120 \pi ^{2}  \hbar \mu _{0 \chi } ^{2}  }    \left( 3 B _{\chi i } B _{\chi j } - \delta _{ij}  B _{\chi } ^{2}   \right)  f _{-2} ( \Lambda _{\chi} )  . \label{Sigma3_fin}
\end{align} 
Summing up the three contributions,  (\ref{Sigma1_fin}), (\ref{Sigma2_fin}) and (\ref{Sigma3_fin}), we establishes the result of Eq. (\ref{Conductivity_Tensor_MagMoment_Final}) for the orbital magnetic moment contribution $\sigma _{ij} ^{(\chi , m )} ({\boldsymbol{B}} _{\chi},T)$.

\bibliography{PHE.bib}
\end{document}